\documentclass[sigconf, nonacm]{acmart}

\usepackage{amsmath,amsfonts}
  
\usepackage{amsthm,amsmath,amssymb}
\usepackage{mathrsfs}
\usepackage{array}

\usepackage[linesnumbered,ruled,noline,norelsize]{algorithm2e}

\usepackage[capitalise]{cleveref}
\usepackage{array}
\usepackage[caption=false,font=scriptsize,labelfont=sf,textfont=sf,margin=0pt,parskip=0pt,farskip=0pt,captionskip=1pt]{subfig}

\usepackage{stfloats}
\usepackage{url}
\usepackage{verbatim}

\usepackage{graphicx}
\usepackage{float}

\usepackage{bbding}

\usepackage{colortbl}
\usepackage[cmyk]{xcolor}
\usepackage{multirow}
\newcolumntype{T}{>{\tiny}l} 
\newcolumntype{S}{>{\small}l} 
\newcolumntype{H}{>{\Huge}l} 
\newcolumntype{F}{>{\footnotesize}l} 
\newcolumntype{Z}{>{\scriptsize}l} 

\newcommand{\nosemic}{\renewcommand{\@endalgocfline}{\relax}}
\newcommand{\dosemic}{\renewcommand{\@endalgocfline}{\algocf@endline}}
\let\oldnl\nl
\newcommand{\nonl}{\renewcommand{\nl}{\let\nl\oldnl}}

\usepackage[straightquotes]{newtxtt}

\newcommand\vldbdoi{XX.XX/XXX.XX}
\newcommand\vldbpages{XXX-XXX}
\newcommand\vldbvolume{19}
\newcommand\vldbissue{1}
\newcommand\vldbyear{2026}
\newcommand\vldbauthors{\authors}
\newcommand\vldbtitle{\shorttitle} 
\newcommand\vldbavailabilityurl{URL_TO_YOUR_ARTIFACTS}
\newcommand\vldbpagestyle{plain}

\begin{document}
\title{PAT: Pattern-Perceptive Transformer for Error Detection in Relational Databases}

\author{Jian~Fu}
\affiliation{%
  \institution{Harbin Institute of Technology, China}
}
\email{fujian@stu.hit.edu.cn}

\author{Xixian~Han}
\affiliation{%
  \institution{Harbin Institute of Technology, China}
}
\email{hanxx@hit.edu.cn}

\author{Xiaolong~Wan}
\affiliation{%
  \institution{Harbin Institute of Technology, China}
}
\email{wxl@hit.edu.cn}

\author{Wenjian~Wang}
\affiliation{%
  \institution{Shanxi University, China}
}
\email{wjwang@sxu.edu.cn}

\begin{abstract}
Error detection in relational databases is critical for maintaining data quality and is fundamental to tasks such as data cleaning and assessment.
Current error detection studies mostly employ the multi-detector approach to handle heterogeneous attributes in databases, incurring high costs.
Additionally, their data preprocessing strategies fail to leverage the variable-length characteristic of data sequences, resulting in reduced accuracy.
In this paper, we propose an attribute-wise PAttern-perceptive Transformer (PAT) framework for error detection in relational databases. 
First, PAT introduces a learned pattern module that captures attribute-specific data distributions through learned embeddings during model training. 
Second, the Quasi-Tokens Arrangement (QTA) tokenizer is designed to divide the cell sequence based on its length and word types, and then generate the word-adaptive data tokens, meanwhile providing compact hyperparameters to ensure efficiency. 
By interleaving data tokens with the attribute-specific pattern tokens, PAT jointly learns shared data features across different attributes and pattern features that are distinguishable and unique in each specified attribute. 
Third, PAT visualizes the attention map to interpret its error detection mechanism.
Extensive experiments show that PAT achieves excellent F1 scores compared to state-of-the-art data error detection methods. Moreover, PAT significantly reduces the model parameters and FLOPs when applying the compact QTA tokenizer.

\end{abstract}

\maketitle

\pagestyle{\vldbpagestyle}
\begingroup\small\noindent\raggedright\textbf{PVLDB Reference Format:}\\
\vldbauthors. \vldbtitle. PVLDB, \vldbvolume(\vldbissue): \vldbpages, \vldbyear.\\
\href{https://doi.org/\vldbdoi}{doi:\vldbdoi}
\endgroup
\begingroup
\renewcommand\thefootnote{}\footnote{\noindent
This work is licensed under the Creative Commons BY-NC-ND 4.0 International License. Visit \url{https://creativecommons.org/licenses/by-nc-nd/4.0/} to view a copy of this license. For any use beyond those covered by this license, obtain permission by emailing \href{mailto:info@vldb.org}{info@vldb.org}. Copyright is held by the owner/author(s). Publication rights licensed to the VLDB Endowment. \\
\raggedright Proceedings of the VLDB Endowment, Vol. \vldbvolume, No. \vldbissue\ %
ISSN 2150-8097. \\
\href{https://doi.org/\vldbdoi}{doi:\vldbdoi} \\
}\addtocounter{footnote}{-1}\endgroup

\ifdefempty{\vldbavailabilityurl}{}{
\vspace{.3cm}
\begingroup\small\noindent\raggedright\textbf{PVLDB Artifact Availability:}\\
The source code, data have been made available at \url{https://github.com/big-datalab/PAT}.
\endgroup
}

\section{Introduction}
Error detection is crucial for improving data quality in relational databases, enabling reliable analytics and decision-making in the tasks of production scheduling~\cite{data_cleaning2019}, medical diagnosis~\cite{fan2012foundations}, and financial auditing~\cite{nashaat2021tabreformer}. 
For instance, analyzing supermarket datasets enables optimized product placement strategies to increase sales revenue. However, such datasets often suffer from missing price entries and inconsistent category labels due to manual input errors or system integration issues, necessitating error detection to ensure reliable insights~\cite{ganti2022data}. 
Similarly, healthcare databases used for clinical predictions must detect errors like mismatched medical codes or duplicate records to prevent biased recommendations~\cite{fan2012foundations}.

Relational databases produced in real-world applications typically exhibit diverse data error types, such as the following well-known error types: typo mistakes~\cite{conf/sigmod/MahdaviAFMOS019}, missing values~\cite{CleanML2021}, outliers~\cite{abedjan2016detecting,CleanML2021} and attribute domain violations~\cite{conf/naacl/IidaTMI21}. For example, $\mathcal{D}$ is a restaurant database as shown in Table~\ref{tab:tableexample}, including three tuples $t_1, t_2$ and $t_3$ which are related to three attributes, i.e. \texttt{ID}, \texttt{Name} and \texttt{Address}. \texttt{ID} refers to a registration number ranging from one to five-digit integers. \texttt{Name} represents the restaurant. \texttt{Address} represents a specific location with long texts. Hence, `1xx19' in $t_2$ is a typo mistake caused by the erroneously inserting, omitting, or substituting of characters. $t_2$ has a missing value mistake because the value of its \texttt{Address} is empty. `Closed' in $t_2$ is an error of attribute domain violation since it shows the status of the business rather than the required restaurant name;
`6e-2' of $t_3$ is an outlier as it is the fractional number and deviates from the expected range.

\begin{table}[t]
        \caption{A restaurant database $\mathcal{D}$.}
        \label{tab:tableexample}
        \centering
        \begin{tabular}{cccp{4.2cm}}
        \toprule
        \textbf{TID}  & \textbf{ID}   & \textbf{Name}   &\textbf{Address}   \\
        \midrule
        $t_1$ & 10018 & COFFEA  & 3621 N   Western Ave … Capital City, Province 12345\\
        $t_2$ & 1xx19 & Closed  & \\
        $t_3$ & 6e-2  & BAKERY          & Condo Complex Name,   Building B, Unit 12 … Hometown, FL 34102\\
        \bottomrule
        \end{tabular}
        \end{table}

Motivated by the importance of detecting errors in relational databases, there are numerous works proposed in literature~\cite{journals/ase/CoteNAHK24,journals/pvldb/NiMZWLY24}. 
Rule-based algorithms use statistics~\cite{chu2016data} and predefined constraints~\cite{conf/edbt/rein} to detect the single error type such as outlier~\cite{pit2016outlier}, violations of attribute dependencies~\cite{dallachiesa2013nadeef}.
However, these algorithms are restricted in detecting confined error types and hard to further improve their performance~\cite{chu2015katara}.

Detecting multiple types of data errors simultaneously remains a significant challenge. 
Traditional approaches, such as rule-based~\cite{dallachiesa2013nadeef} and predefined constraints methods~\cite{conf/sigmod/WangH19}, incur high costs and complexity, as they require domain expertise and iterative validation~\cite{data_cleaning2019,journals/debu/NeutatzCA021}. 
These methods also struggle to adapt to heterogeneous error types and scale effectively in various datasets.

Thus, researchers currently focus on addressing such problems with machine learning (ML) techniques~\cite{krishnan2017boostclean,pham2021spade,CleanML2021}, because they can reduce reliance on predefined rules by automatically detecting data errors. Even though certain ML methods require extensive feature constructing and lack robustness for complex errors and multi-attribute domains~\cite{krishnan2016activeclean} on relatively large-scale databases~\cite{journals/pvldb/NiMZWLY24}, the approaches based on deep learning (DL)~\cite{conf/sigmod/HoloDetect19, conf/sigmod/Miao0021} perform good. 

DL model provides vast parameters to comprehensively exploit the features of multi-attribute structures and diverse error types. 
Nowadays, many studies design effective sequence models~\cite{conf/sigmod/Miao0021, liu2022picket, conf/edbt/HolzerS22, peng2022self} to enable end-to-end detection of diverse data errors in the databases.
Existing DL-based error detection methods can be categorized into three types based on the model architecture and the data granularity being detected, which are \textit{multi-detector cell-level approaches}~\cite{conf/sigmod/MahdaviAFMOS019, conf/sigmod/HoloDetect19, neutatz2019ed2, pham2021spade, abdelaal2024saged}, \textit{single-model cell-level approaches}~\cite{conf/sigmod/Miao0021, conf/edbt/HolzerS22}, and \textit{large-model tuple-level approaches}~\cite{liu2022picket, conf/naacl/IidaTMI21, nashaat2021tabreformer, conf/icde/zeroed}.
The first employs multiple detectors, and each detector detects data errors for cells in the specified attribute domain to achieve high accuracy effectively; 
the second utilizes a single model to detect data errors for cells across multiple attributes in databases; 
the third takes a tuple as input and trains the large model to detect data errors in all attribute values in the tuple~\cite{conf/naacl/IidaTMI21,nashaat2021tabreformer,liu2022picket}. 
Besides, the existing DL-based methods in their data preprocessing stage typically use the tokenizer~\cite{conf/sigmod/Miao0021, liu2022picket} in which cell sequence is broken down as pieces named tokens, and then they are mapped to token embeddings that can be used for a model.

However, in this paper, the existing DL-based error detection methods are limited in the following three aspects.

\textit{First, inappropriate tokenization exists in database preprocessing.}
As the preprocessing techniques in error detection, tokenizing and embedding operations should be more concerned with the morphological representation of data in relational databases, such as token composition and format factors (word types, varying text lengths etc.). 
Existing tokenizers~\cite{bojanowski2017enriching,conf/acl/SennrichHB16a, conf/icassp/SchusterN12} follow natural language processing (NLP) conventions and their tokenization criteria concentrate on semantic features.
In databases, specifically,
most cells contain short to medium-length sequences, with attributes typically being numerical or categorical~\cite{liu2022picket}. 
A smaller portion of cells, however, contain longer sequences, which are often found in specific attributes, e.g. film description, and full cast in the movies~\cite{neutatz2019ed2}.

\textit{Second, methods require high accuracy but with low cost.}
As for efficiency, when detecting data errors using DL techniques, 
multi-detector approaches will employ a large number of classifiers for databases with numerous attributes,
and large-model tuple-level approaches entail increased parameters and computation.
Consequently, both approaches incur higher training and deployment costs.
To reduce time and space cost, model architecture necessitates simplicity and integrity.
In terms of accuracy, single-model approaches struggle to learn the distributions of multi-attribute domains, as each attribute has distinct data profiles and error types.
Such inter-attribute biases~\cite{journals/air/VilaltaD02} inevitably diminish accuracy and generalization.

\textit{Third, error detection models are short of explainability.}
As an infrastructure for data quality management, the error detection model must be interpretable for its internal mechanism. 
When processing unstructured data (e.g., textual paragraphs~\cite{vaswani2018tensor2tensorneuralmachinetranslation, vig-2019-multiscale} and image patches~\cite{dosovitskiy2021animage, He_2021_ICCV}),
DL models leverage attention mechanisms to offer rich contextual information for visualization and analysis. 
However, DL-based error detection in databases provides fewer attention parameters, making it challenging to establish explainable relationships between input sequences and classification outputs.

To address these challenges, we propose an attribute-wise PAttern-perceptive Transformer framework (PAT) to detect data errors in relational databases.
In the preprocessing phase, 
this paper introduces the quasi-tokens arrangement (QTA) algorithm that divides variable-length data sequences into data tokens to achieve database tokenization. 
It is found that long cell sequences take up a small proportion in databases, whereas leading to token redundancy and high computation costs.
To address it, a compact QTA tokenizer which is fit for most short sequences is provided to retain model efficiency and scalability.
In the PAT architecture, 
a learned pattern module is incorporated, where learned pattern tokens provide distinguishable features among multiple attributes, 
which are combined with the data tokens in an alternate order as the interleaved tokens.
By processing the interleaved tokens as input, the PAT network disentangles the attribute-specific patterns from the shared data features of all cells in the database.
These attribute-specific patterns are encoded in dedicated pattern tokens and pattern-wise parameters, while the shared data features are stored in data-wise parameters within the PAT network.
By integrating the QTA tokenizer and the pattern module, PAT breaks the trade-offs between accuracy and efficiency.
Moreover, to address the interpretability challenge, PAT visualizes the attention scores to construct the relations between input data and prediction output.
By analyzing attention maps, data and pattern tokens serve different roles in error detection, 
and erroneous tokens can be inferred among data tokens.
Furthermore, the erroneous words in the input sequence can be used for error locating and repairing in relational databases.

We summarize our contributions as follows:

1. A novel framework. We propose an attribute-specific \underline{{\textbf{PA}}}ttern-perceptive \underline{{\textbf{T}}}ransformer framework (PAT) that interleaves the data tokens with learned pattern tokens jointly as input to facilitate learning features of data values and the corresponding attribute profiles in the PAT network.

2. Data preprocessing. A \underline{{\textbf{Q}}}uasi-\underline{{\textbf{T}}}okens \underline{{\textbf{A}}}rrangement (QTA) tokenizer is introduced to separate the cell sequence into data tokens based on appropriate length and word types, supporting both default and compact modes. 

3. Model interpretability. We extract attention scores from the PAT network to visualize the relationship between input data and classification predictions during detecting errors.

4. Experimental evaluation.
Extensive experiments demonstrate that our PAT method achieves superior accuracy and efficiency compared to existing data error detection methods.

\section{RELATED WORK}

We review the existing error detection methods from the perspective of model architecture and data levels being detected, 
including statistics, rules and ML based data error detection approach, single-model cell-level data error detection approach, multi-classifier cell-level data error detection approach, large-model tuple-level data error detection approach, and learned pattern module.

\textit{Statistics, Rules, and ML based Approach.}
Traditional error detection methods rely on predefined rules and constraints to detect individual or multiple error types.
NADEEF~\cite{dallachiesa2013nadeef} use defined rules to detect the violated cell values.
\textsc{dBoost} \cite{pit2016outlier} use statistics to detect outliers.
KATARA \cite{chu2015katara} utilizes knowledge bases to detect semantic violations.
To detect more error types,
ensemble methods~\cite{visengeriyeva2018metadata} combine multiple statistical techniques for distinct error types into a unified framework.
Since rule-based methods~\cite{journals/tkde/DingWSWLG22} have better performance on limited error types, 
recently there are works~\cite{conf/sigmod/BaoBBDFLLLLLOTW24, journals/pacmmod/PirhadiMCMS24} that combine rules and ML models to broadly detect more kinds of errors.
Therefore, the above methods require collecting extensive rules and constraints and manual parameter tuning, which fall into lack of efficiency.

\textit{Single-model Cell-level Approach.} 
ETSB-RNN~\cite{conf/edbt/HolzerS22} utilizes the bidirectional RNN model with shallow network layers to detect cells in databases. 
As for data preprocessing in error detection, tokenization procedures for a cell sequence contain data tokenizing and embedding.
The tokenizer regards each character \cite{conf/edbt/HolzerS22}, word (tokenize a cell by whitespace~\cite{conf/sigmod/Miao0021,liu2022picket}), or cell~\cite{ nashaat2021tabreformer} as a token.
As for data embedding,
characters-to-numerics mapping (customized index mapping) is used as data embedding in ETSB-RNN. 
Besides the above character embedding manner, the more often way to process the cell value is word embedding including Word2Vec model~\cite{NIPS2013_mikolov} e.g. fastText, Gensim, spaCy, which learns relationships in sequences automatically and brings more semantic information.
So far, methods that preprocess data in tokenization still follow the routine of NLP and do not make full use of the formation and structure of relational databases.

\textit{Multi-detector Cell-level Approach.} 
Raha~\cite{conf/sigmod/MahdaviAFMOS019}, ED2~\cite{neutatz2019ed2} and HoloDetect~\cite{conf/sigmod/HoloDetect19} use multi-detectors use attribute-wise multi-detectors to train each classifier for each attribute.
HoloDetect uses data augmentation to expand the pool of erroneous training samples,
and extracts rich predefined features for the classifier model.
As a drawback, these two techniques collectively increase memory consumption when processing large-scale datasets.
In the data preprocessing stage,
HoloDetect and Picket~\cite{liu2022picket} use word representations model fastText~\cite{bojanowski2017enriching} to perform token-to-vector (tok2vec) process.
Except for tokenization,
ED2~\cite{neutatz2019ed2}, METADATA~\cite{visengeriyeva2018metadata} and SPADE~\cite{pham2021spade} take the predefined aggregated multi-hierarchical features as input, which capture more concerned and context-dependent information to make prediction.

\textit{Large-model Tuple-level Approach.}
ActiveClean~\cite{krishnan2016activeclean} proposes a concept where all attributes of a tuple are simultaneously detected and got classification output. 
Works~\cite{conf/sigmod/Miao0021, wang2023sudowoodo} that treat data cleaning as a sequence classification task, which converts a cell $d_{i,j}$ or a tuple $t_{i}$ into the format "[COL] $a_1$ [VAL] $d_{i,j}$" or "[COL] $a_1$ [VAL] $d_{i,1}$ ... [COL] $a_m$ [VAL] $d_{i,m}$", respectively. But to bring the attribute name into input is not enough to represent the features of the responding attribute.
ROTOM~\cite{conf/sigmod/Miao0021} uses the pre-trained large language model (LLM) Roberta to detect dirty data. Moreover, ZeroED~\cite{conf/icde/zeroed} leverage the LLM reasoning ability to assist error detection.
TabReformer~\cite{nashaat2021tabreformer} focuses on the errors of inter-tuples and inter-attributes in databases. 
Tabbie~\cite{conf/naacl/IidaTMI21} employs the pre-trained LLM ELECTRA to construct structural relationships of tuples and attributes.
However, applying LLM in error detection consumes plenty of time and computing resources.



\begin{figure*}[htbp] 
        \centering
        \includegraphics[width=0.98\textwidth]{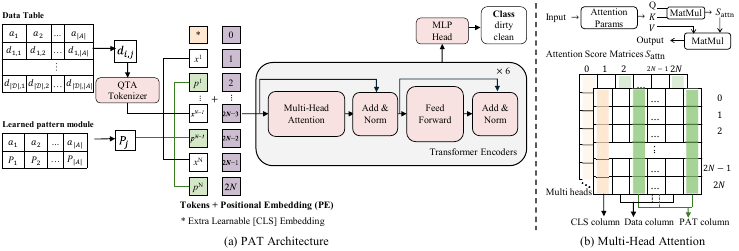}
        \caption{Overview of proposed Pattern-Perceptive Transformer (PAT) model architecture. 
        } 
        \label{PATmodel}
        \end{figure*}


\section{PRELIMINARIES}
Error detection (EDT) in relational databases is defined as 
identifying violations of predefined formats and rules constrained by relational databases~\cite{abedjan2016detecting}.
Let a relational database \( \mathcal{D} \) consist of tuples \( t_i \) (\( i$$\in$$\{1, \ldots, |\mathcal{D}|\} \)) with schema \( A = \{a_1, \ldots, a_{|A|}\} \), where \( d_{i,j} \) denotes the value of attribute \( a_j \) in tuple \( t_i \).  
A data error occurs when \( d_{i,j} \) is inconsistent with the true observation or the ground truth value \( d_{i,j}^* \in \mathcal{D}^* \), i.e.,  
\[
\text{ErrorData} = \left\{d_{(i,j)}|d_{(i,j)}\neq d_{(i,j)}^*, d_{(i,j)}\in \mathcal{D}, d_{(i,j)}^*\in \mathcal{D}^*\right\},
\]  
where \( \mathcal{D}^* \) represents the ground truth database. 

Errors in relational databases are categorized into syntactic errors (cell-level errors) and semantic errors (tuple-level errors).
Syntactic errors include  
outliers (OT)~\cite{abedjan2016detecting,CleanML2021}, 
missing values (MV)~\cite{CleanML2021}, 
formatting violations (FV)~\cite{visengeriyeva2018metadata}, 
typo mistakes (TP)~\cite{conf/sigmod/MahdaviAFMOS019}.
FV and TP belong to pattern violations (PV)~\cite{abedjan2016detecting}
Semantic errors can be further classified into the inter-row level errors which have Duplicates (DP)~\cite{conf/sigmod/WangH19} and the inter-column level errors which contain violations of functional dependencies (VFD)~\cite{journals/tkde/WanHWL24, conf/sigmod/BaoBBDFLLLLLOTW24} and attribute domain violations (AD)~\cite{conf/naacl/IidaTMI21}.
Our work focuses on detecting a broad range of syntactic errors, and also detecting specific semantic errors, i.e. Attribute Domain violations (AD). 

\section{PAT FRAMEWORK}

To deal with the data error detection problem, we develop an attribute-specific pattern-perceptive Transformer (PAT) framework.
To further enrich the feature representation of cells in databases, we propose a quasi-token arrangement (QTA) algorithm (Section \uppercase\expandafter{\romannumeral4}.A). 
PAT model incorporates a learned pattern module to learn features in both data and their attribute patterns (Section \uppercase\expandafter{\romannumeral4}.B).
We also integrate an explainability measure for the error detection model (Section \uppercase\expandafter{\romannumeral4}.C). 
Finally, we analyze the PAT framework in terms of attention scores as well as the time and space complexity of the QTA algorithm (Section \uppercase\expandafter{\romannumeral4}.D).

\begin{algorithm}[!t]
        \KwIn{Database $\mathcal{D}$, Default/compact mode $m$} 
	\KwOut{Prediction Output $O_{\mathrm{pred}}$} 
        \DontPrintSemicolon
	hyperparameters $D, N$ $\leftarrow$ TokenDimNumSetup($\mathcal{D}$, $m$)  \label{algoPAT-line:hyperparameters}\\
        data embeddings $T_{\mathrm{train}}, T_{\mathrm{infer}}$$\leftarrow$QTA$-$Tokenizer($\mathcal{D}$, $D, N$) \label{algoPAT-line:tokenizer} \\
        Initialize PAT Model $M$ with $D, N$ \\
        Initialize pattern embeddings $P_{1,...,|A|}$ for dataset $\mathcal{D}$ \\

        \textbf{repeat} \\
                \quad load data $T_{i,j}$, attribute id $a[j]$ from $T_{\mathrm{train}}$ \label{algoPAT-line:forstart} \\
                \quad $M$ $\leftarrow$ PAT$_{\mathrm{training}}$($T_{i,j}$, $P_{a_j}$) \\
        \textbf{until convergence} \label{algoPAT-line:forend} \\
        
        $O_{\mathrm{pred}}$ $\leftarrow$ $M$$(T_{\mathrm{infer}})$// PAT model inference

	\caption{PAT Model Learning Algorithm}
	\label{alg:PAT}
\end{algorithm}

PAT framework unifies the above three components in an end-to-end fashion.
As shown in Fig.~\ref{PATmodel}(a), the data tokens and the pattern tokens are combined in an interleaved form as input tokens in the PAT framework. 
We further analyse procedures in Algorithm~\ref{alg:PAT}.
First, to prepare the data tokens,
in the data preprocessing stage, we propose a QTA tokenizer algorithm to break down the cell sequence $d_{i,j}$ with inconstant-length into a set of data tokens $T_{d_{i,j}}$ and create their embedding tokens $T_E$ with fixed token dimension $D$ and number $N$ (line~\ref{algoPAT-line:tokenizer}).
Specifically, $D$ and $N$ are computed in the tokenization hyperparameters setup algorithm(line~\ref{algoPAT-line:hyperparameters}).
Second, 
to fully utilize the Transformer architecture~\cite{NIPS2017_attention,dosovitskiy2021animage} on relational databases,
and observing that each attribute in databases has different appearances and characteristics,
PAT incorporates the learned attribute-wise pattern module into the Transformer encoder-based classification network~\cite{dosovitskiy2021animage}.
Thus,
PAT model $M$ and learned pattern tokens $P_{j}$ for the attribute $a_j$ are obtained when PAT training(lines~\ref{algoPAT-line:forstart}-\ref{algoPAT-line:forend}).
Third, there is a potential demand to inspect the inner mechanism of the designed PAT model to enhance model transparency and reveal why and how the DL model can detect data errors.
As illustrated in Fig.~\ref{PATmodel}(b), we extract the attention score from MSA as the attention map during the model inferencing to interpret error detection processes.

\subsection{Database Tokenization}

\subsubsection{QTA Tokenizer} 

        \begin{figure*}[htbp] 
        \centering 
        \includegraphics[width=0.98\textwidth]{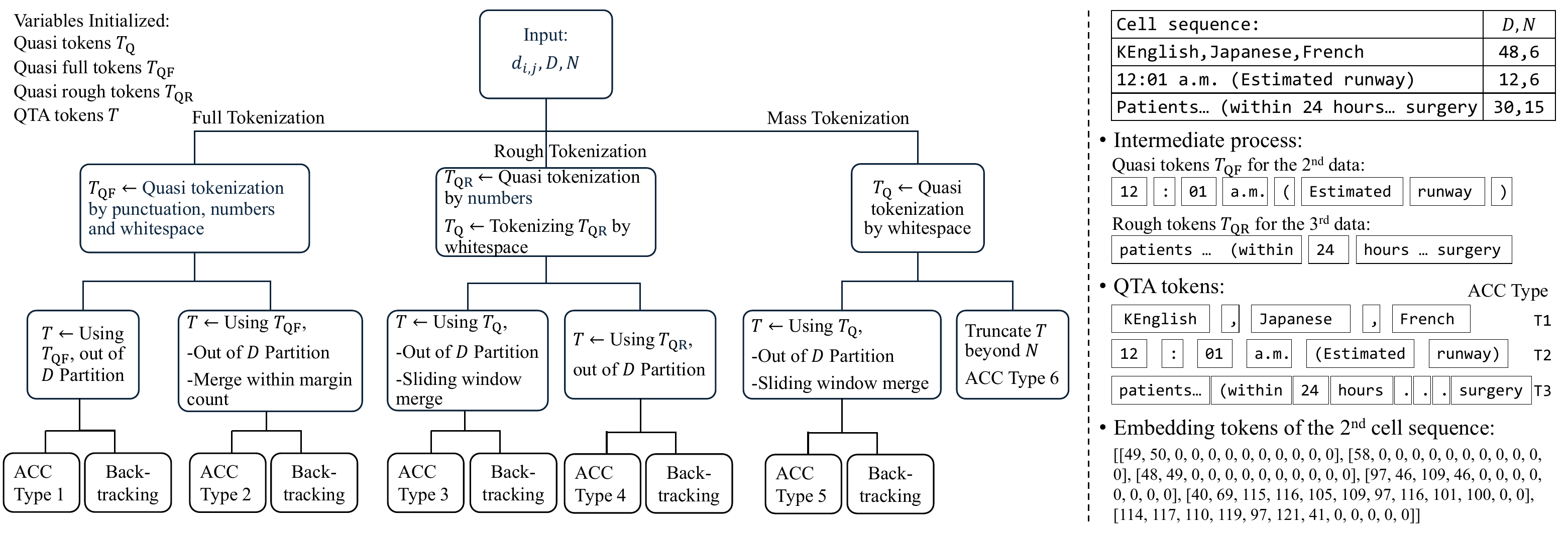} 
        \caption{Demonstrate QTA Tokenizer with Tree-like Backtracking.}
        \label{QTA_Illustration}
        \end{figure*}

        In the data preprocessing stage, Quasi-Token Arrangement (QTA) algorithm is proposed to segment the cell sequence as quasi tokens by quasi tokenizers and further orchestrate them to obtain data tokens for the EDT task.
        QTA algorithm can be named shortly as QTA tokenization and QTA tokenizer.
        To handle variable-length sequences and meet the restrictions on token dimension $D$ and numbers $N$, 
        the QTA algorithm partition input sequence with three procedures depending on delicate to rough tokenizing degree: (1) full tokenization, (2) rough tokenization with numerics separated, and (3) mass tokenization without numerics separated, which correspond to three branches in the backtracking tree in Fig.~\ref{QTA_Illustration}.

        To put concretely, the three procedures have general routines including quasi tokenization and arrangement measures,
        which are implemented in different steps depending on the declared separators and division degree.
        From the perspective of composition, the text sequence can be regarded as the combination of numbers, words and punctuation marks. 
        The first step of the QTA algorithm is to apply the quasi-tokenizer in raw sequences which refers to the tree nodes of depth 1 in Fig.~\ref{QTA_Illustration} and get quasi-tokens as processed sources to be arranged.
        Subsequently, for each token out of $D$ in quasi-tokens $T_{\mathrm{Q}}$, we partition it into several tokens.
        For quasi-tokens who do not meet the requirements, full tokenization merges adjacent tokens under the conditions that words and punctuation marks are given priority when combining (with a higher priority level), while numbers are considered last (with a lower priority level). Additionally, tokens in the same level are merged first.
        The above arrangement measures including partitioning and merging are operated in the backtracking tree nodes of depth 2 as illustrated in Fig.~\ref{QTA_Illustration}.
        Generally speaking, full tokenization is suitable for processing short data sequences, while rough tokenization and mass tokenization are applicable for handling long sequences as well as ultra-long sequences.

\begin{algorithm}[!t]
	Initialize: Unicode character mapping $\mathcal{M}$ \\
	\tcp{\begin{small}Full Tokenization w/ punct num independence\end{small}}
	$T_{\mathrm{QF}} \leftarrow$ QF-Tokenizing $d_{i,j}$ by punct num whitespace \label{algo1-line:full_start}\\ 
	
	$T_{\mathrm{Q}} \leftarrow$ OutDimPartition($T_{\mathrm{QF}}$), $\,\,$$N_{\mathrm{m}} \leftarrow N-$ $N_{\mathrm{Q}}$ \label{algo1-line:full_branch2}\\
	\textit{If $N_{\mathrm{Q}}$$<$$N$ then $T_{\mathrm{acc}} \leftarrow  T_{\mathrm{Q}}$ and \textbf{goto} Line~\ref{line:1}}\hfill // T1 \label{algo1-line:T1}\\
	$N_{\mathrm{p}} \leftarrow$ The size of punctuation tokens in $T_{\mathrm{Q}}$ \\
	\textit{If $N_{\mathrm{m}}$$\leq$$N_{\mathrm{p}}$ then $T_\mathrm{acc}$$\gets$}MG($T_{\mathrm{Q}}$) \textit{and \textbf{goto} Line~\ref{line:1}} \hfill//T2\label{algo1-line:T2} \label{algo1-line:full_end}\\
	\tcp{\begin{small}Rough Tokenization with num independence\end{small}}
	$T_{\mathrm{QR}} \leftarrow$ Q-Tokenizing $d_{i,j}$ by numerics \label{algo1-line:rough_start}\\
	$T_{\mathrm{Q}} \leftarrow$ Tokenize each $t$ in $ T_{\mathrm{QR}}$ by blank  \label{algo1-line:rough_T3_start}\\
	$T_{\mathrm{Q}} \leftarrow$ OutDimPartition($T_{\mathrm{Q}}$) \\
	$T_{\mathrm{}} \leftarrow$ Merge $T_{\mathrm{Q}}$ with num independence \\
	\textit{If $ N_{\mathrm{T}} < N$ then $T_{\mathrm{acc}} \leftarrow  T_{\mathrm{}}$ and \textbf{goto} Line~\ref{line:1}} \hfill// T3 \label{algo1-line:rough_T3}\\
	
	$T_{\mathrm{}2} \leftarrow$ OutDimPartition($T_{\mathrm{QR}}$) \label{algo1-line:rough_T4_start}\\

	\textit{If $N_{\mathrm{T}2} < N$ then $T_{\mathrm{acc}} \leftarrow  T_{\mathrm{}2}$  and \textbf{goto} Line~\ref{line:1}} \hfill// T4 \label{algo1-line:rough_T4_end}\\
	
	\tcp{\begin{small}Mass Tokenization without priority\end{small}}
	$T_{\mathrm{Q}} \leftarrow$ Q-Tokenizing $d_{i,j}$ by whitespace \label{algo1-line:mass}\\
	$T_{\mathrm{Q}} \leftarrow$ OutDimPartition($T_{\mathrm{Q}}$) \label{algo1-line:mass_start}\\ 
	$T_{\mathrm{QTA}} \leftarrow$ Merge $T_{\mathrm{Q}}$ without independent tokens \\
	
	$T_{\mathrm{acc}}\gets\,$TokensUpdate$(T_{\mathrm{QTA}})$\hfill // T5 and T6 \label{algo1-line:mass_end}\\
	
	Embeddings  $E \leftarrow$ $\mathcal{M}$($T_{\mathrm{acc}}$) // Unicode index map \label{line:1}\\ 
	$T_E \leftarrow $ padding zero to each vector $e$ in $E$ \label{algo1-line:end}\\ 
	\textbf{return} data token embeddings $T_E$  \\
	
	\caption{QTA-Tokenizer($d_{i,j}$, $D$, $N$)}
	\label{alg:QTA}
\end{algorithm}

        In Algorithm~\ref{alg:QTA},
        QTA tokenizer algorithm takes as input the cell value $d_{i,j}$, where $i\in [1,\ldots, \left| \mathcal{D} \right|],$ $j \in$ $[1,$ $\ldots,$ $ \left| A \right|]$, along with the token dimension $D$, and the numbers of tokens $N$.
        In the three procedures, the tokens will be accepted if meeting the token dimension $D$ and number $N$ requirements.

        Full tokenization: Quasi-Full-tokenizer (QF-tokenizer) divides cell value $d_{i,j}$ by whitespace (as separator), numbers and punctuation to obtain the quasi tokens $T_{\mathrm{QF}}$ (Line~\ref{algo1-line:full_start}). 
        After partitioning the out of dimension tokens, the tokens $T_{\mathrm{Q}}$ will be accepted as type 1  (Lines~\ref{algo1-line:full_branch2}-\ref{algo1-line:T1}) if qualify the token number $N$ restrictions.
        Otherwise,
        if the count of punctuation tokens $N_{\mathrm{p}}$ is equal or greater than margin count $N_{\mathrm{m}}$ (the difference value between $N$ and quasi-tokens count $N_{\mathrm{Q}}$),
        full tokenization will merge the punctuation token into the adjacent word token within margin count as MG ($T_{\mathrm{Q}}$) and return the tokens that are accepted as type 2 (Line~\ref{algo1-line:T2}). 
        As for the unsatisfied occasions in which cell values are usually long sequences, the QTA tokenizer will backtrack and conduct the subsequent tokenization procedures in rough and mass mode.

        Rough tokenization:  Quasi-rough tokens $T_{\mathrm{QR}}$ come from the way of quasi-tokenizer with numerics separators (not considering punctuation marks and whitespace) (Line~\ref{algo1-line:rough_start}).
        In backtracking branch of type 3 in Fig.~\ref{QTA_Illustration}, it is necessary to use whitespace tokenizer to divide $T_{\mathrm{QR}}$ as $T_{\mathrm{Q}}$ and get the quasi-tokens $T_{\mathrm{Q}}$ partitioned and merged with numbers independence (Lines~\ref{algo1-line:rough_T3_start}-\ref{algo1-line:rough_T3}).
        If going into the branch of type 4, the rough tokenizer will partition the token if out of $D$ in quasi-rough tokens $T_{\mathrm{QR}}$ and obtain $T$. If meeting the acceptance restrictions, tokens $T$ will be returned as $T_{\mathrm{acc}}$ (Lines~\ref{algo1-line:rough_T4_start}-\ref{algo1-line:rough_T4_end}).

        Mass tokenization: For extra-long sequences, the third procedure does not consider the priority of any token types. Therefore, quasi-tokenizer segments cell value $d_{i,j}$ by whitespace separator and regards them as quasi-tokens $T_{\mathrm{Q}}$ (Line~\ref{algo1-line:mass}).
        Subsequently, the backtracking  branch of type 5 has the same partitioning and merging procedures as that of type 3 and acquires $T_{\mathrm{QTA}}$.
        Eventually, in TokensUpdate() will accept $T_{\mathrm{QTA}}$ as type 5 when it qualifies the $N$ conditions, otherwise the function will truncate the tokens $T$ beyond $N$ and accept it as type 6 (Lines~\ref{algo1-line:mass_start}-\ref{algo1-line:mass_end}).

        In the final step of the QTA algorithm, data token embeddings are generated through a character index mapping mechanism (Lines~\ref{line:1}-\ref{algo1-line:end}).
        Specifically, QTA uses Unicode character indices to transform $T_{\mathrm{acc}}= \{ \tau_1,...,\tau_N\}$ into data token embeddings $T_E=\{x_{}^{i}|x_{}^{i}\in {{\mathbb{R}}^{{D}}};i=1,2,\cdots ,N\}$, 
        which preserve the original structure and morphological features of the cell sequence,
        as the examples illustrated in Fig.~\ref{QTA_Illustration}.
        Data token embeddings $T_{E}$ from the QTA tokenizer can represent all the morphological characteristics of cell sequence strictly, which is beneficial to detect most cell-level data errors in the PAT model.
        Also, the well-arranged tokens by delicately tokenizing reserve the intact formation of words and phrases, and help to localize erroneous words precisely, consequently providing foundations for repairing the error values.
        
        Practically in the data preprocessing stage, the QTA tokenizer precomputes all cells from $\mathcal{D}$, constructing two critical components: the data token embedding vocabulary $V_{\mathrm{tokens}}$ for all tokens and cell value dictionary $Dict_{\mathrm{cells}}$ for all cells in a database $\mathcal{D}$.
        The precomputed structures enable efficient retrieval of token embeddings for any specified  cell sequence $d_{i,j}$
        during the training and inferencing of the PAT model, directly accessing $Dict_{\mathrm{QTA}}$ to avoid redundant computation.

\subsubsection{Tokenization Hyperparameters Settlement}

PAT provides auto-configured hyperparameters to adapt the QTA tokenization algorithm for various databases and make the trade-off between the accuracy and efficiency of error detection.
As described in Algorithm~\ref{alg:hyperparamsset}, the tokenization hyperparameters setup algorithm and its sub-procedure that is Critical Point Find (CPF) are devised to determine the token dimension $D$ and the number of tokens $N$ with both the default and compact mode, thereby realizing the word-adaptive QTA tokenization.

\begin{algorithm}[!t]
        
        \caption{TokenDimNumSetup(database $\mathcal{D}$)}
        \SetKwInput{KwInput}{Input}                
        \SetKwInput{KwOutput}{Output}              
        \DontPrintSemicolon
        
        \SetKwFunction{FSub}{\textnormal{CriticalPointFind}} 
                 \label{alg:hyperparamsset}
       Coefficients and their compact version $\beta$, $\beta _\mathrm{c}$ \\ Initialize boundary ratio $\mu_{\mathrm{l}}$, $\mu_{\mathrm{cl}}$, $\mu_{\mathrm{r}}$, $\mu_{\mathrm{cr}}$, $\mu_{\mathrm{long}}$ \\
                      \textbf{for} each cell value $d_{i,j}$ in $\mathcal{D}$ \textbf{do} \label{algo2-line:forstart}\\
                          \quad $T_{i,j} \leftarrow$ QF-Tokenizer($d_{i,j}$), Update($V_{\mathrm{QF}}$, $T_{i,j}$)\\
                          \quad $S$.append(size of $T_{i,j}$)  \\
                          \quad $L_{\mathrm{c}}$.append(length of $d_{i,j}$)  \\
                      \textbf{end} \label{algo2-line:forend}\\
                          
                      $L_{\mathrm{QF}} \leftarrow$  $[$ len($\tau$) for each token $\tau$ in $V_{\mathrm{QF}}$ $]$ \label{algo2-line:LQF}\\
          
                      \tcp{Hyperparameters setting in default mode}
                      $D \leftarrow$ CriticalPointFind($L_{\mathrm{QF}}$, $\beta, \mu_{l}$, $\mu_{\mathrm{r}}$) \label{algo2-line:HPdefault}\\ 
                      $N \leftarrow$ CriticalPointFind($S$, $\beta, \mu_{l}$, $\mu_{\mathrm{r}}$) \label{algo2-line:HPdefaultend}\\
          
                      \tcp{Hyperparameters setting in compact mode}
                      $D_{\mathrm{c}} \leftarrow$ CriticalPointFind($L_{\mathrm{QF}}$, $\beta_\mathrm{c}$, $\mu_{\mathrm{cl}}, \mu_{\mathrm{cr}}$) \label{algo2-line:HPcompact}\\
                      $s_{\mathrm{long}} \leftarrow$  Percentile$(L_{\mathrm{c}}, \mu_{\mathrm{long}})$,   
                      $N_{\mathrm{c}} \leftarrow$ $\lfloor s_{\mathrm{long}}/D_{\mathrm{c}}\rfloor$ \label{algo2-line:HPcompactend}\\

              \KwRet $D,N,D_{\mathrm{c}},N_{\mathrm{c}}$\; 
              
              \nonl\noindent\rule[0.25\baselineskip]{0.45\textwidth}{0.5pt} \\ 
              \SetKwProg{Fn}{Procedure}{}{}
              \Fn{\FSub{$\mathrm{array}$$\,a,\beta,\mu$\begin{tiny}$_{\mathrm{left}}$\end{tiny}$,\mu$\begin{tiny}$_{\mathrm{right}}$\end{tiny}}}{
                  $y_{\mathrm{type}} \leftarrow [ ]$,
                  $y_{\mathrm{H}}, x_{\mathrm{H}} \leftarrow$ FreqHist($a$, $N_{\mathrm{bins}}$) \label{algo2-line:FH}\\

                  $p_{\mathrm{left}}, p_{\mathrm{right}} \leftarrow$  Percentile$(a, \mu_{\mathrm{left}}, \mu_{\mathrm{right}})$ \label{algo2-line:p-lr} 
                  
                  \textbf{for} each $p_{\mathrm{value}}$ in Range ($p_{\mathrm{left}}$,$p_{\mathrm{right}}$) \textbf{do} \label{algo2-line:range}\\
                
                  \quad \uIf {$y_{\mathrm{H}}[\mathrm{ArrayIdx}(x_{\mathrm{H}}, p_{\mathrm{value}})] < \mathrm{Aver}(y_{\mathrm{H}})\cdot \beta$\label{algo2-line:ifstart}}{
                          \qquad $y_{\mathrm{type}}$.append(label= "low") \\
                  }
                  \quad \uElse{ \qquad $y_{\mathrm{type}}$.append(label="high")} 
                  \quad \textbf{end} \label{algo2-line:ifend}\\
                  \textbf{end}

                  \KwRet $x_{\mathrm{H}}[$BinaryDividePointIdx($y_{\mathrm{type}}$)$]$\; \label{algo2-line:end}} 

        \end{algorithm}

In Algorithm~\ref{alg:hyperparamsset},
the hyperparameters of the QTA tokenizer including token dimension $D$ and token number $N$ are provided in the default and compact modes to meet the accuracy and efficiency in PAT.
First, taking raw database $\mathcal{D}$ as input and 
applying the full quasi-tokenizer to each cell in the dirty databases, we update $T_{i,j}$ on the vocabulary set $V_{\mathrm{QF}}$ as Update($V_{\mathrm{QF}}$, $T_{i,j}$), the size list $S$ of data tokens $T_{i,j}$ for all cells and length list of cells $L_{\mathrm{c}}$ (Lines~\ref{algo2-line:forstart}-\ref{algo2-line:forend}). 
Then the length list $L_{\mathrm{QF}}$ contains the length of each token $t$ in the vocabulary set $V_{QF}$ (Line~\ref{algo2-line:LQF}).
In the default and compact mode, the token dimension $D$  and $D_{\mathrm{c}}$ are computed in the CPF procedure using $L_{\mathrm{QF}}$ as the common argument, 
while the other arguments have default ($\beta$,$\mu_{\mathrm{l}}$,$\mu_{\mathrm{r}}$) and compact ($\beta_{\mathrm{c}}$,$\mu_{\mathrm{cl}}$,$\mu_{\mathrm{cr}}$) version (Lines~\ref{algo2-line:HPdefault},\ref{algo2-line:HPcompact}). 
Typically, the coefficient $\beta_{\mathrm{c}}$ in compact mode is higher than the one $\beta$ in default mode whose values depend on different databases, as a higher $\beta_{\mathrm{c}}$ prioritizes high-frequency points, which correspond to shorter dimensions.
The token numbers $N$ is determined in the CPF procedure with the argument $S$ (Line~\ref{algo2-line:HPdefaultend}).
Since in the compact mode PAT framework has to reduce the computation,
the estimated length of long sequence $s_{\mathrm{long}}$ is the $\mu_{\mathrm{long}}$-th percentile of the cell length list $L_\mathrm{c}$ in dataset $\mathcal{D}$ where the argument $\mu_{\mathrm{long}}$ aims to drop the rare long cell sequences.
Thus, the fitting numbers of the tokens $N_{\mathrm{c}}$ is calculated by  $s_{\mathrm{long}}$  dividing $D_{\mathrm{c}}$ (Line~\ref{algo2-line:HPcompactend}). 
Finally, the algorithm returns the tokenization hyperparameters $D$, $N$, $D_{\mathrm{c}}$ and $N_{\mathrm{c}}$.

In the Critical Point Find (CPF) procedure in Algorithm~\ref{alg:hyperparamsset}, 
we make frequency histogram (FreqHist) with array $a$ and bin count $N_{\mathrm{bins}}$ as arguments to obtain list of points $x_{\mathrm{H}}$ and their frequency values $y_{\mathrm{H}}$ (Line~\ref{algo2-line:FH}). 
Before selecting the critical point in array $a$, we use percentile() with ratio arguments $\mu_{\mathrm{left}}$ and $\mu_{\mathrm{right}}$ to get the $\mu_{\mathrm{left}}$-th and $\mu_{\mathrm{right}}$-th percentile of the array $a$ that is $p_{\mathrm{left}}$ and $p_{\mathrm{right}}$.
Therefore, we use $p_{\mathrm{left}}$ and $p_{\mathrm{right}}$ as two end points in array $a$ to narrow the range of interest (ROI) and find the critical point (Lines~\ref{algo2-line:p-lr}-\ref{algo2-line:range}).
Then CPF averages $y_{\mathrm{H}}$ and multiplies it with $\beta$ as referential frequency threshold $\beta\cdot$Aver($y_{\mathrm{H}}$). 
For the index $\mathrm{ArrayIdx}(x_{\mathrm{H}})$ of each point in ROI, CPF procedure gets its frequency value in $y_{\mathrm{H}}$ and compares it with referential threshold.
If lower than the referential threshold, the frequency value is classified into the low-frequency bar, otherwise, the high-frequency bar. So the frequency bars in ROI get the type list $y_{\mathrm{type}}$ with the high and low types (Lines~\ref{algo2-line:ifstart}-\ref{algo2-line:ifend}). 
The key step of CPF procedure is to find the index to divide the type list $y_{\mathrm{type}}$ into two segments which represents as BinaryDividePointIdx($y_{\mathrm{type}}$) and return the critical point in $x_{\mathrm{H}}$ (Line~\ref{algo2-line:end}).

QTA tokenizer with compact hyperparameters provides less number of tokens and short token dimension to accommodate all the text pieces which is especially suitable for the short and medium cell sequences, and gets rough or mass tokens for long sequences to reduce the time and space cost and maximize detecting speed in the EDT task.
PAT with default QTA hyperparameters will make the delicate arrangement for the long sequences and gain higher accuracy on the EDT task.

\subsection{Pattern-Perceptive Transformer Architecture}

\subsubsection{Learnable Attribute-wise Patterns}

It can be observed that values of each attribute in a relational table share a feature distribution, and this correspondence can be viewed as analogous to the relationship between a value and its projection.
Therefore, we arrange the data tokens to be detected $T_{d_{i,j}}$$=$$\{x^k | k$$=$$1,2,...,N\}$ and pattern tokens $P$$=$$\{ p^k_{j} | k$$=$$1,...,N; j$$=$$1,...,|A|\}$ in an interleaved order, 
where $N$ is the number of tokens, and each token $x^k$ is an embedding vector with dimension $D$.
The interleaved tokens represent the paired and complementary relationship between a cell sequence and its attribute domain.
In addition, a learnable embedding of CLS token $x_{\ell=0}^{0} = x_{\mathrm{cls}}$ is prepended to the interleaved tokens, whose state at the output of the Transformer encoder ($Z_{\ell}^{0}$) acts as classification feature representation $y$. 
Integrating a CLS token, data tokens and pattern tokens, the input tokens can be presented as follows:

\begin{equation}{{{Z}}_{0}}=[{{x}_{\mathrm{cls}}}; x_{}^{1}; p^{1}; x_{}^{2}; p^{2}; \cdots ; x_{}^{N}; p^{N}]+{{E}}\end{equation}
Where ${Z}_{0},{E} \in {{\mathbb{R}}^{(1+2N)\times D}}$. ${E}$ provides the extended position information for the interleaved data and pattern tokens. The positonal arrangement is the difference between our interleaved layout in PAT and the separated layout in \cite{zheng2022template}.

Specifically, during the training process, cells in the attribute domain $a_j$ in the dataset collectively learn  $N$ pattern embedding tokens independently which contains differential data and error features unique to the attribute $a_j$. 
Therefore, the pattern embedding tokens are shared within the same attribute domain and independent in different attribute domains.

As presented in Fig.~\ref{PATmodel}, PAT model has a stack of $\mathcal{L}=6$ layers of the encoder and attaches MLP head to $Z_{\mathcal{L}}^0$ to make the classification.
In Equation \eqref{eq:pat}, the encoder in PAT consists of multi-head self-attention (MSA) and multi-layer perceptron (MLP) blocks. Besides, LayerNorm (LN) and residual connections are applied. 
The $\ell{th}$ encoder layer outputs the feature embeddings ${Z}_{\ell}\in {{\mathbb{R}}^{(1+2N)\times D}}$.
Finally, PAT outputs the classification as $y$.
\begin{align} {Z}_{\ell}^{^\prime} &= \text{MSA}(\text{LN}({Z}_{\ell-1}^{}))+{Z}_{\ell-1}^{},\ell=1,...,\mathcal{L}
        \notag
        \\ {Z}_{\ell} &= \text{MLP}(\text{LN}({Z}_{\ell}^{^\prime}))+{Z}_{\ell}^{^\prime},\ell=1,...,\mathcal{L} 
        \notag
        \\ y &=\text{MLPHead}(Z_{\mathcal{L}}^0)
        \label{eq:pat}
\end{align}

\subsubsection{Self-Attention Mechanism}

As the core network structure in the PAT model, self-attention effectively constructs the relations between the error detection output and the cell sequence with its attribute-wise patterns.
The self-attention function processes the interleaved token embeddings (${Z}_{\ell}$) as input. 
In the first step, the input embeddings are linearly projected into queries, keys and values matrices noted as $Q$, $K$, $V$.
The projection process is 
$\label{eq_qkv} [Q,K,V] = \text{LN}({Z}_{\ell}\cdot [W_Q,W_K,W_V])$
where $W_Q,W_K,W_V\in {{\mathbb{R}}^{D\times d_k}} $ are parameter matrices of MSA structure, and $Q,K,V\in {{\mathbb{R}}^{(1+2N)\times d_k}}$.

The self-attention function $\label{eq_SA} \text{SA}({Z}_{\ell})=\text{softmax}(\frac{Q{{K}^{\text{T}}}}{\sqrt{{{d}_{k}}}})V$ is computed parallelly with $Q$, $K$, $V$ and output the attention matrix. The dimension of keys is denoted by ${d}_{k}$.

To further analyze the internal structure of self-attention, we decompose it into separate steps. 
First, we formally define the attention score matrix as ${{S}_{\text{attn}}}=\operatorname{Softmax}(\frac{Q\cdot{{K}^{\text{T}}}}{\sqrt{{{d}_{k}}}})$.
To neatly illustrate the constituent elements of the attention scores, we retrieve $Q\cdot{{K}^{\text{T}}}$ as $S_{QK}$, defined in Equation (\ref{eq_qk}) (The CLS elements are ignored).
$Q$ and $K$ can be described as $Q=[q_{{{x}_{\mathrm{cls}}}}^{{}}; q_{{{x}^{1}}}^{{}}; q_{{{p}^{1}}}^{{}}; \cdots; q_{{{x}^{N}}}^{{}}; q_{{{p}^{N}}}^{{}}]$, 
$K=[k_{{{x}_{\mathrm{cls}}}}^{{}}; k_{{{x}^{1}}}^{{}}; k_{{{p}^{1}}}^{{}}; \cdots; k_{{{x}^{N}}}^{{}}; k_{{{p}^{N}}}^{{}}]$.
Among the element vectors in $Q$, a selected example $q_{{{x}_{\mathrm{cls}}}}$  represents the query value corresponding to the CLS token. The element vectors in $K$ and $V$ do the same way.

\begin{equation}  \label{eq_qk}
        \begin{aligned}
     \begin{split}
        & {{S}_{QK}}={Q\cdot{{K}^{\text{T}}}}  \\ 
        & =\left[ {\footnotesize \begin{matrix} 
        q_{{{x}^{1}}}\cdot k_{{{x}^{1}}}^{\text{T}} & q_{{{x}^{1}}}^{{}}\cdot k_{{{p}^{1}}}^{\text{T}} & \cdots  & q_{{{x}^{1}}}^{{}}\cdot k_{{{x}^{N}}}^{\text{T}} & q_{{{x}^{1}}}^{{}}\cdot k_{{{p}^{N}}}^{\text{T}}  \\
        q_{{{p}^{1}}}\cdot k_{{{x}^{1}}}^{\text{T}} & q_{{{p}^{1}}}\cdot k_{{{p}^{1}}}^{\text{T}} & \cdots  & q_{{{p}^{1}}}^{{}}\cdot k_{{{x}^{N}}}^{\text{T}} & q_{{{p}^{1}}}^{{}}\cdot k_{{{p}^{N}}}^{\text{T}}  \\
        {\vdots} & {\vdots} & \ddots  & {\vdots} & {\vdots}  \\
        q_{{{x}^{N}}}^{{}}\cdot k_{{{x}^{1}}}^{\text{T}} & q_{{{x}^{N}}}^{{}}\cdot k_{{{p}^{1}}}^{\text{T}} & \cdots  & q_{{{x}^{N}}}^{{}}\cdot k_{{{x}^{N}}}^{\text{T}} & q_{{{x}^{N}}}^{{}}\cdot k_{{{p}^{N}}}^{\text{T}}  \\
        q_{{{p}^{N}}}^{{}}\cdot k_{{{x}^{1}}}^{\text{T}} & q_{{{p}^{N}}}^{{}}\cdot k_{{{p}^{1}}}^{\text{T}} & \cdots  & q_{{{p}^{N}}}^{{}}\cdot k_{{{x}^{N}}}^{\text{T}} & q_{{{p}^{N}}}^{{}}\cdot k_{{{p}^{N}}}^{\text{T}}  \\
\end{matrix}} \right]  
\end{split}
\end{aligned}
\end{equation}

Taking the data tokens and pattern tokens in an alternate format as input,
the internal attention matrix (${S}_{QK}$), as presented in Equation~\eqref{eq_qk}, combines and fuses the data and pattern elements.
In the PAT network, data-wise parameters learn the indistinguishable features shared across different attribute domains, which can facilitate learning errors that occur frequently (e.g. MV, TP) and sequence format (i.e. common word types and length).
On the other hand, pattern parameters extract the distinguishable and unique features that are dependent on the specified attribute, consequently enhancing the recognition ability of exclusive errors (e.g. OT error in ID, AD error in Name) and sequence format (e.g. attributes that have long sequences).
Overall, the learned pattern features are externally independent among different attribute domains and internally shared within the same attribute domain.

Finally, we compute the output matrix of self-attention: 
$\text{SA}({Z}_{\ell})={{S}_{\text{attn}}}\cdot {{\left[ \begin{matrix}
        \begin{matrix}
                v_{{{x}_{\mathrm{cls}}}} & v_{{{x}^{1}}} & v_{{{p}^{1}}}  \\
     \end{matrix} & \cdots  & v_{{{x}^{N}}} & v_{{{p}^{N}}} \\
     \end{matrix} \right]}}
$ 
where the value matrix $V$ can be described as $ V$$=$$[v_{{{x}_{\mathrm{cls}}}}; v_{{{x}^{1}}}; v_{{{p}^{1}}}; \cdots; v_{{{x}^{N}}}; v_{{{p}^{N}}}]  $.

In the training phase, data samples with different attributes are randomly selected for training, while their associated pattern tokens dynamically switch accordingly. 
Thus, the pattern embeddings in the PAT model act as bridges for multi-attribute data interacting in databases, enabling the error detection model to perform multi-attribute detection. 
Considering the capability in the semantic comprehension of the cell text sequences and 
the effects of the interaction among attributes via pattern embeddings, 
PAT can detect Attribute Domain violations (AD), which is a semantic error. 

\subsection{ Inspecting and Interpreting Pattern-Perceptive Transformer}

The explainability module selects the last layer of the 6 stacked encoder layers in the PAT model since the latter layer learns the more features concerned with error classification.
Because of the multi-head ($h$$=$8) parallel structure, the explainability module retrieves a total of 8 matrices of attention scores.
As for the classification of the EDT task,
different from the NLP models whose output embeddings are parallel to input tokens, PAT appoints a CLS embedding component to build the connection between the CLS state at the encoder output and cell tokens at the input.
In the next step,
as shown in Fig.~\ref{PATmodel}(b), we retrieve the first column vector $S_{\mathrm{vis}}$ (an orange color) from each attention score matrix $S_{\text{attn}}$.
\begin{equation}S_{\mathrm{vis}}=S_{\text{attn}}[:h][:N][1]\cdot \eta \label{eq:svis} \end{equation}
As presented in Equation~\eqref{eq:svis},
the explainability mechanism uses attention scores $S_{\mathrm{vis}}$ to visualize the relationship between $N$ input tokens and CLS prediction output (the first embedding).
When taking the retrieved attention vector to visualize,  
the attention lines are too shallow to represent the inner relationship for most data sequences. 
Hence, we multiply the retrieved attention scores $S_{\text{attn}}$ by coefficient $\eta $ to highlight the line depth. 
Moreover, we remove the unnecessary rear tokens corresponding to zero-padding features generated in Algorithm~\ref{alg:QTA} when token numbers are less than $N$.

\subsection{Framework Analysis}

\subsubsection{QTA Algorithm Analysis}

\textit{Notations and their inequalities:}
The time and space complexity of QTA depends on token dimension $D$ and number $N$, the length of input sequence $L$, 
numerical tokens count $N_{nums}$, quasi-tokens count $n$ which is short for $N_{\mathrm{Q}}$, 
average length for quasi-tokens $L_{\mathrm{avg}}$. 
To help analyze the algorithm, quasi-tokens are divided into two categories ($l\leqslant D$ and $l>D$) which depend on whether the length of quasi-token $l$ is greater than $D$ or not.
Hence, the average length for tokens of the two categories is $L_{l\leqslant D}$ and $L_{l>D}$. 
The count for two categories is $N_{l\leqslant D}$ and $N_{l>D}$, and we agree $N_{l\leqslant D}\gg N_{l>D}$ on all relational databases on massive data.
The equation comes that $L=L_{l\leqslant D}N_{l\leqslant D}+L_{l>D}N_{l>D}=L_{\mathrm{avg}}(N_{l\leqslant D}+N_{l>D})$ and $n=\frac{L}{L_{\mathrm{avg}}}$. 
On massive data, we assume that input tokens have the capacity of cell sequence i.e. $DN \geqslant L$.
Thus, we draw the inequality relations that for length variable $L_{l>D}>D>\frac{L}{N}>L_{\mathrm{avg}}$, 
for count variable $n>N_{l\leqslant D}>N>\frac{L}{D}>N_{l>D}$ 
and for sequence length $DN>L>L_{\mathrm{avg}}N>L_{l\leqslant D}N_{l\leqslant D}>L_{l>D}N_{l>D}$.

\textit{Time complexity:}
In Fig.~\ref{QTA_Illustration}, the conditions of type 2, 3, and 5 have a nested loop, i.e. quasi tokenizer and arrangement measures (partitioning and merging).
Specifically for type 3, rough tokenizer is incorporated into quasi tokenizer i.e. quasi-tokens count $n$ contains numerical tokens count $N_{nums}$.
Therefore, these 3 types have higher time and space complexity than other types, e.g. type 1 or 4.
The time cost of the outer loop (quasi tokenizer)  is $N_{l\leqslant D}+N_{l>D}$ i.e. $\frac{L}{L_{\mathrm{avg}}}$.
The time cost of the inner loop (partitioning and merging) is $\frac{D}{N_{l\leqslant D}}$ for merging or $\frac{L_{l>D}}{D}$ for partitioning. 
Note that merging and partitioning do not operate on numerical tokens for the type 2, 3, and 4.
Overall, the time cost is $N_{l\leqslant D}\frac{D}{L_{l\leqslant D}}+N_{l>D}\frac{L_{l>D}}{D}$.
For the first item, 
due to the inequality $\frac{L}{L_{\mathrm{avg}}}>N_{l\leqslant D}$ and $L_{l\leqslant D}>\frac{L_{\mathrm{avg}}(L-{{L}_{\mathrm{avg}}}N)}{L}$ ($\frac{L}{L_{\mathrm{avg}}}L_{l\leqslant D}>L-L_{l>D}N_{l>D}$), 
we can derive $N_{l\leqslant D}\frac{D}{L_{l\leqslant D}}<\frac{L^2D}{L_{\mathrm{avg}}^2(L-{{L}_{\mathrm{avg}}}N)}<\frac{n^{2}D}{L-{{L}_{\mathrm{avg}}}N}$
For the second item, since the inequality $DN>L_{l>D}N_{l>D}$, we can derive that $N_{l>D}\frac{L_{l>D}}{D}<N$.
Since $n>N$, the first item has a higher order of growth.
Therefore, the time complexity of QTA algorithm is $\mathcal{O}(\frac{D\cdot n^{2}}{L-{{L}_{\mathrm{avg}}}N})$.

\textit{Space complexity:} The space cost of the outer loop is $\frac{L}{L_{\mathrm{avg}}}$, and that of the inner loop is $\frac{L}{N_{l\leqslant D}}$ for merging and $\frac{N_{l>D}}{D}$ for partitioning.
The space complexity of QTA is $\mathcal{O}({n}+\frac{D\cdot {n}}{L-{{L}_{\mathrm{avg}}}N})$.

Based on the above analysis, the QTA tokenizer algorithm has a quadratic time complexity and a linear space complexity. This performance profile is efficient for the database preprocessing stage.

\section{EXPERIMENTS}
We do comprehensive evaluations on the proposed PAT framework. 

{\bf{Dataset Settings.}}
Table~\ref{tab_datasets_profile} shows the statistics on dataset sizes, vocabulary sizes, error types and error rates in the datasets.
Among all datasets, Adult, Beers, Flights, Food, Hospital, Movies, Rayyan, Restaurants, Soccer are real-world datasets, 
while Billionaire\cite{conf/icde/zeroed}, HOSP-100K~\cite{arocena2015messing} and Tax~\cite{arocena2015messing} are synthetic datasets whose errors were generated by error generation engines.
As for error types,
these datasets cover various types of errors including most cell-level errors, 
inter-row level errors (DP in Hospital) and inter-column level errors (AD in Beers).
Flights have mostly departure or arrival time errors that violate inter-column dependencies (VAD) and MV errors. 
The Hospital, Rayyan, and Soccer datasets serve to identify error detection methods under conditions of low error proportions.
Moreover, we list the count of numbers and words in the vocabulary generated from our QF-Tokenizer.
Adult, Beers, Food, Restaurants and Tax have more numerical cells. 
Flights, Hospital, HOSP-100K have number and character mixed cells, and especially Restaurants contains attribute values made up of words, numerics and punctuation marks, including attributes like website, street address, extra multiple phone, etc.
Scalability for attribute values is assessed using long-sequence datasets (Hospital, Movies, Rayyan, Restaurants), while scalability for tuples is evaluated with three large-scale datasets (HOSP-100K, Restaurants, Tax).

\begin{table}[]
        \centering
        \arrayrulecolor{black}
        \caption{Experimental data sets}
        \label{tab_datasets_profile}
        \setlength{\tabcolsep}{0.1cm}
        \begin{tabular}{p{2.2cm}p{1.3cm}p{0.58cm}p{1.3cm}l}
                \toprule
                Data Set   & \begin{tabular}[c]{@{}l@{}}Size (Row \\ and Col.)\end{tabular} & \begin{tabular}[c]{@{}l@{}}Error\\ Rate\end{tabular} & \begin{tabular}[c]{@{}l@{}}Vocabulary \\ Size ($\times10^3$) \end{tabular} & Error Types     \\
                \hline
                Adult~\cite{liu2022picket}     & (32561,15)                                                      & 0.32                                                 & 2.1, 22                                                                         & OT, VAD         \\
                Beers~\cite{conf/sigmod/MahdaviAFMOS019}     & (2410,11)                                                       & 0.15                                                 & 3.1, 2.9                                                                           & MV, TP, FV, VAD \\
                Billionaire~\cite{conf/icde/zeroed}     & (2615,22)                                                       & 0.09                                                 & 6.8, 0.9                                                                           & MV, PV, OT, VAD \\
                Flights~\cite{conf/sigmod/MahdaviAFMOS019}   & (2376,7)                                                        & 0.30                                                  & 0.1, 2.5                                                                            & MV, TP, FV, AD \\
                Food~\cite{conf/sigmod/HoloDetect19}      & (1900,13)                                                       & 0.06                              & 2.8, 5.5                                                                           & MV, TP, FV           \\
                Hospital~\cite{conf/sigmod/MahdaviAFMOS019}  & (1000,20)                                                      & 0.03                                                 & 1.7, 2                                                                           & TP, FV, VAD, DP   \\
                HOSP-100K~\cite{visengeriyeva2018metadata}  & (100000,17)                                                      & 0.11                                                 & 7.9, 42.5                                                                           & TP, FV, VAD     \\
                Movies~\cite{neutatz2019ed2}     & (7390,17)                                                      & 0.01                                                 & 126.3, 9.6                                                                         & MV, FV          \\
                Rayyan~\cite{conf/sigmod/MahdaviAFMOS019}    & (1000,11)                                                      & 0.08                                                 & 11.6, 3.1                                                                                     & MV, TP, FV, VAD \\
                Restaurants~\cite{nashaat2021tabreformer} & (28787,16)                                                      & 0.01                                                 & 45.1, 39.6                                                                         & OT, MV, TP, FV           \\
                Soccer~\cite{conf/sigmod/HoloDetect19}    & (3879,10)                                                     & 0.06                                                & 0.7, 0.1                                                                              & MV, OT, VAD     \\
                Tax~\cite{conf/sigmod/MahdaviAFMOS019}       & (200000,15)                                                    & 0.04                                                 & 33.9, 49.3                                                                         & TP, FV, VAD     \\ 
                \bottomrule
        \end{tabular}
        \end{table}

\begin{table}[]
        \centering
        \arrayrulecolor{black}
        \caption{Performance Comparison with State-of-the-Art Methods}
        \label{tab_exp_compare}
        \begin{tabular}{p{0.63cm}p{0.13cm}p{0.43cm}p{0.4cm}p{0.79cm}p{0.43cm}p{0.40cm}p{0.40cm}p{0.45cm}l}
        \toprule
        \begin{tabular}[c]{@{}l@{}}Data\\     Sets \end{tabular}                                                                   & \begin{tabular}[c]{@{}l@{}}M\\      \end{tabular} & Raha & ED2   & \begin{tabular}[c]{@{}l@{}}HOLOD\\      -ETECT\end{tabular} & \begin{tabular}[c]{@{}l@{}}ETSB\\      -RNN\end{tabular} & \begin{tabular}[c]{@{}l@{}}ROT\\      -OM\end{tabular}& \begin{tabular}[c]{@{}l@{}}Zero\\      -ED\end{tabular} & PATC & PAT   \\ \hline
        \multirow{3}{*}{Adult}                                                     & P       & 81.3 & 99.1  & \textbf{100}                                                      & \textbf{100}                                                    & -     & - & \textbf{100}       & \textbf{100} \\
                                                                                   & R       & 43.5 & 53.1  & 86.7                                                       & 98.5                                                     & -     & - & 99.8        & \textbf{99.9}  \\
                                                                                   & F1      & 56.7 & 69.1  & 92.9                                                       & 99.3                                                     & -     & - & \textbf{99.9}        & \textbf{99.9}  \\ \hline
        \multirow{3}{*}{Beers}                                                     & P       & 99.0 & \textbf{100} & 88.5                                                       & \textbf{100}                                                    & \textbf{100} & 88.8 & \textbf{100}       & \textbf{100} \\
                                                                                   & R       & 99.0 & 96.0  & 78.6                                                       & 96.0                                                     & 91.9  & 68.9 & \textbf{100}       & \textbf{100} \\
                                                                                   & F1      & 99.0 & 98.0  & 83.3                                                       & 98.0                                                     & 95.8  & 77.4 & \textbf{100}       & \textbf{100} \\ \hline
        \multirow{3}{*}{\begin{tabular}[c]{@{}l@{}}Billion\\      -aire\end{tabular}}    & P       & 71.6   & 46.3    & 35.8                                                        & 70.5                                                       & --    & 72.4 & 99.3          & \textbf{99.8} \\
                                                                                   & R       & 62.6   & 62.9    &  \textbf{87.1}                                                        & 57.3                                                       & --    & 81.2 & 70.8          & 71.2 \\
                                                                                   & F1      & 66.8   & 53.4    & 50.6                                                        & 62.8                                                       & --    & 76.5 & 82.7          & \textbf{83.1} \\ \hline
        \multirow{3}{*}{Flights}                                                   & P       & \textbf{82.0} & 79.0  & 41.7                                                       & 81.0                                                     & -     & 58.6 & 75.8        & 73.6  \\
                                                                                   & R       & 81.0 & 63.0  & 79.4                                                       & 68.0                                                     & -     & 72.2 & 90.0        & \textbf{93.3}  \\ 
                                                                                   & F1      & 81.0 & 68.0  & 54.7                                                       & 74.0                                                     & -     & 73.2 & \textbf{82.0}        & \textbf{82.0}  \\ \hline
        \multirow{3}{*}{Food}                                                      & P       & 99.8 & \textbf{100} & 66.8                                                       & 99.7                                                     & -     & 16.1 & 99.4        & \textbf{100} \\
                                                                                   & R       & 78.2 & \textbf{100} & 52.8                                                       & 93.0                                                     & -     & 94.1 & 99.3        & \textbf{100} \\
                                                                                   & F1      & 87.7 & \textbf{100} & 59.0                                                       & 96.2                                                     & -     & 27.6 & 99.4        & \textbf{100} \\ \hline
        \multirow{3}{*}{\begin{tabular}[c]{@{}l@{}}Hosp\\      -ital\end{tabular}}                                                  & P       & 94.0 & 45.0  & 25.1                                                       & 95.3                                                     & -     & 93.6 & \textbf{100}       & \textbf{100} \\
                                                                                   & R       & 59.0 & 29.0  & 68.3                                                       & \textbf{92.5}                                                     & -     & 71.5 & 36.0        & 41.4  \\
                                                                                   & F1      & 72.0 & 33.0  & 36.7                                                       & 92.7                                                     & \textbf{100} & 81.1 & 53.0        & 58.5  \\ \hline
        \multirow{3}{*}{\begin{tabular}[c]{@{}l@{}}HOSP\\      -100K\end{tabular}} & P       & 68.9 & -     & -                                                          & 67.8                                                     & -     & - & 89.1        & \textbf{89.5}  \\
                                                                                   & R       & 85.8 & -     & -                                                          & 14.2                                                     & -     & - & \textbf{87.3}        & 86.8  \\
                                                                                   & F1      & 76.4 & -     & -                                                          & 22.4                                                     & -     & - & \textbf{88.2}        & 88.1  \\ \hline
        \multirow{3}{*}{Movies}                                                    & P       & 55.6 & 93.0  & 6.8                                                        & 59.1        & -     & 76.8 & 95.8        & \textbf{96.7}  \\
                                                                                   & R       & 4.8  & 5.0   & \textbf{71.6}                                                       & 11.3                                                                                   & -     & 76.7 & 61.3        & 62.9  \\
                                                                                   & F1      & 8.9  & 13.0  & 12.4                                                       & 18.3                                                     & 69.6  & 76.7 & 74.7        & \textbf{76.2}  \\ \hline
        \multirow{3}{*}{Rayyan}                                                    & P       & 81.0 & 80.0  & 80.4                                                       & 87.0                                                     & -     & 77.8 & \textbf{96.9}        & 95.3  \\
                                                                                   & R       & 78.0 & 69.0  & 87.7                                                       & 83.0                                                     & -     & 69.2 & \textbf{96.1}        & 95.3  \\
                                                                                   & F1      & 79.0 & 74.0  & 83.9                                                       & 85.0                                                     & 86.4  & 73.2 & \textbf{96.5}        & 95.3  \\ \hline
        \multirow{3}{*}{\begin{tabular}[c]{@{}l@{}}Restau\\      -rants\end{tabular}}                                               & P       & 32.2 & 76.6  & -                                                          & 89.3                                                     & -     & - & 96.7        & \textbf{97.2}  \\
                                                                                   & R       & 32.3 & 11.5  & -                                                          & 13.9                                                     & -     & - & 69.9        & \textbf{75.3}  \\
                                                                                   & F1      & 32.3 & 19.5  & -                                                          & 24.1                                                     & -     & - & 81.1        & \textbf{84.8}  \\ \hline
        \multirow{3}{*}{Soccer}                                                    & P       & 83.2 & 84.5  & 97.3                                                       & 75.2                                                     & -     & - & \textbf{100}       & \textbf{100} \\
                                                                                   & R       & 78.1 & 77.5  & \textbf{100}                                                      & 47.3                                                     & -     & - & \textbf{99.8}        & 99.7  \\
                                                                                   & F1      & 80.6 & 80.8  & 98.6                                                       & 56.6                                                     & -     & - & \textbf{99.9}        & \textbf{99.9}  \\ \hline
        \multirow{3}{*}{Tax}                                                       & P       & 84.1 & \textbf{100} & -                                                          & 82.0                                                     & -     & - & 99.6        & \textbf{100} \\
                                                                                   & R       & 97.9 & 95.4  & -                                                          & 92.0                                                     & -     & - & \textbf{99.3}        & \textbf{99.3} \\
                                                                                   & F1      & 90.5 & 97.6  & -                                                          & 86.0                                                     & 98.7  & - & 99.5        & \textbf{99.7}  \\ 
                                                                                   \bottomrule

\end{tabular}
\end{table}

{\bf{Experimental Settings.}}
The task of identifying erroneous cells in a database is actually a binary classification problem, where erroneous and clean labels correspond to positive and negative, respectively. 
As metrics for measuring Data Transformer training and inference results, we report precision (P) defined as evaluating the proportion of true positives among all the samples predicted as positive, recall (R) defined as the fraction of TP among all the error samples (actual positives); and F1 score is computed considering both of P and R metrics.

Our model is implemented on the PyTorch framework. We run all the experiments on an Ubuntu 20.04 Linux machine with Intel Xeon Gold 6226R CPU, 256GB memory and 4 NVIDIA GeForce RTX 3090 GPUs.
We follow the principle of separating training and testing stages, i.e. the test set is extracted before the training phase and used after the training is completed.
The model is trained with cross-entropy loss and the Adam optimizer. The learning rate (LR) is initialized as 0.002, followed by the cosine learning rate declining.
Since the token dimension $D$ for our datasets falls into the range of 10 to 72, coordinately we set the MLP dimension $d_{\mathrm{MLP}}=4D$ or $5D$ (typically 128 or 256) in MSA. The dimension of query, key and value (${d}_{q}$, ${d}_{k}$, ${d}_{v}$) are all set to 64. The error bars in following experimental figures indicate the standard deviation from 10 repetitions of each experiment. The performance metrics, M (Precision, Recall, F1-score), are presented as percentages and averaged across multiple trial repetitions.

We experiment with the following four variants of Data Transformer (PAT, PATC, DT, DTC), which belong to single-model cell-level approaches.
DataTrans (DT) is the Transformer encoder for database with the default Quasi-Token Arrangement (QTA) tokenizer for the data preprocessing. DataTrans Compact (DTC) applies the compact QTA tokenizer in the data preprocessing stage.
PAttern-perceptive Transformer (PAT) and PAttern-perceptive Transformer Compact (PATC) are enhanced by learned pattern module and QTA algorithm with separately default and compact hyperparameters.

\begin{figure*}[!t]
        \centering
        \subfloat[Rayyan, Syntactic errors]{\includegraphics[width=0.248\textwidth]{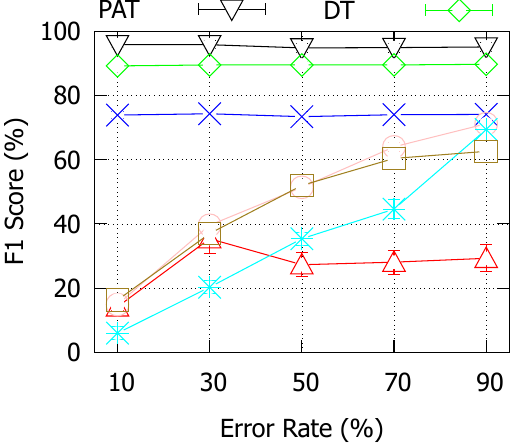}\label{fig:TypeCompare1}}
        \subfloat[Rayyan, Semantic errors]{\includegraphics[width=0.248\textwidth]{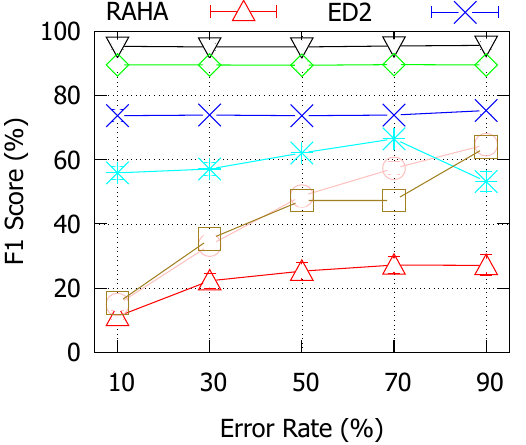}\label{fig:TypeCompare2}}
        \subfloat[Flights, Syntactic errors]{\includegraphics[width=0.248\textwidth]{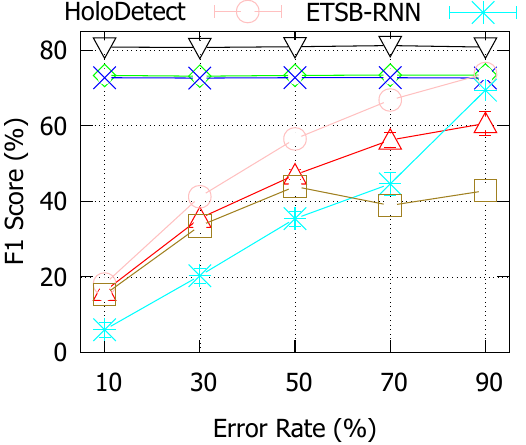}\label{fig:TypeCompare3}}
        \subfloat[Flights, Semantic errors]{\includegraphics[width=0.248\textwidth]{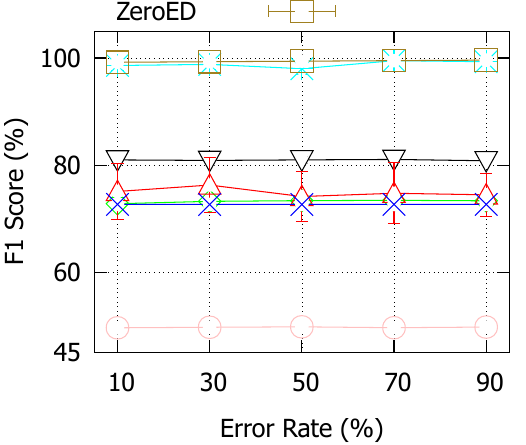}\label{fig:TypeCompare4}}

        \caption{Effectiveness comparison of error detection methods with varying error rate. (Line colors indicate methods.) 
        }
        \label{fig:TypeCompare}
    \end{figure*}

{\bf{Exp 1: Comparison with state-of-the-art methods.}}

As shown in Table~\ref{tab_exp_compare}, our PAT and PATC methods are compared with 5 SOTA methods in P, R and F1 score metrics. The symbol "-" appears when out of memory in HoloDetect and ZeroED, N$\slash$A in ROTOM and ED2.
The evaluation indicates that, as a single-model error detection method, PAT has significant advantages in accuracy, compared to multi-detector approaches (Raha~\cite{conf/sigmod/MahdaviAFMOS019}, ED2~\cite{neutatz2019ed2}, HoloDetect~\cite{conf/sigmod/HoloDetect19}), single-model approaches (ETSB-RNN~\cite{conf/edbt/HolzerS22}) and large-model approaches (ROTOM~\cite{conf/sigmod/Miao0021}, ZeroED~\cite{conf/icde/zeroed}). 
As equipped with the compact QTA tokenizer, PATC has advantages in efficiency and meanwhile gains highest accuracy in Beers, Flights, HOSP-100K, Rayyan, Soccer datasets.
In Flights and Rayyan datasets, PAT and PATC manage to detect tough error types (OT, PV), thereby making solid improvements. 
Overall, our PAT and PATC methods consistently outperform all SOTA methods across datasets except Hospital.
In Hospital, PAT, PATC, and most EDT methods get low accuracy, except ETSB-RNN and ROTOM, which are ascribed to two causes.
First, these methods fail to detect the TP errors involving inserting or substituting character 'x', because it is hard to distinguish such errors in the tokenized cell sequence through character or word embedding.
Second, the Hospital contains many duplicated tuples, and erroneous samples make up only a small proportion of the training data. This leads to imbalanced learning for EDT models as they struggle to generalize from the skewed data distribution.

In Table~\ref{tab_exp_compare},
As a large-model approach, ROTOM is inferior in detecting most cell-level errors i.e. formatting and syntactic violations (FV, TP) in Beers, Movies, Rayyan and Tax.
It is ascribed that its LLMs are not adaptive in the EDT task and cannot take advantage of the structural database,
though it brings powerful representation for words on context and semantic expression.
ETSB-RNN takes character index mapping in its preprocessing stage and assigns each character in the input sequence to each input neuron in the model, thereby effectively detecting the character insertion and substitution error in Hospital. 
HoloDetect gains high accuracy in datasets with low error rates, such as Beers and Rayyan.
For large-scale datasets, to refrain from out-of-memory issues when running HoloDetect, we remove the data augmentation module and only use part of the dataset to extract features.
As for large-scale datasets, HoloDetect runs through Hospital and Movies, and fails to manage Restaurants and Tax.
Lastly, all the baselines show accuracy decline in Flights because they are hard to detect outliers (OT) that deviate from real numeric values.

\begin{figure}[!t]
        \centering
        \subfloat[Restaurants]{\includegraphics[width=0.248\textwidth]{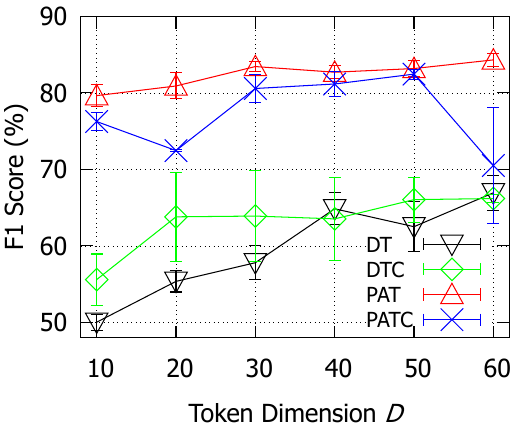}\label{fig:DwithNcompare1}}
        \subfloat[Rayyan]{\includegraphics[width=0.248\textwidth]{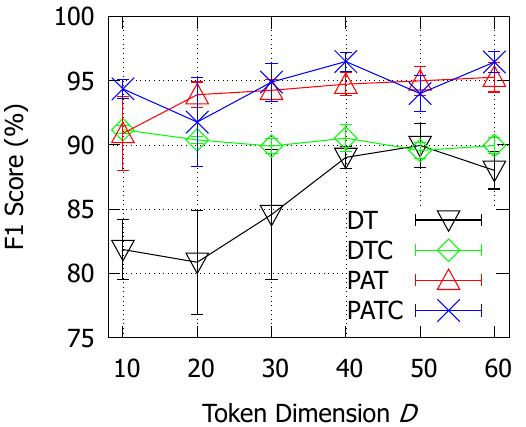}\label{fig:DwithNcompare2}}
        \\
        \subfloat[Movies]{\includegraphics[width=0.248\textwidth]{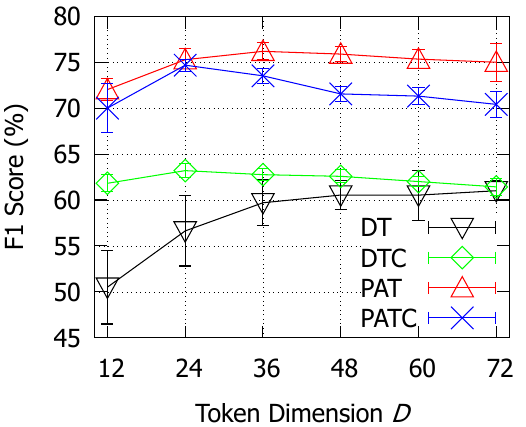}\label{fig:DwithNcompare3}}
        \subfloat[HOSP-100K]{\includegraphics[width=0.248\textwidth]{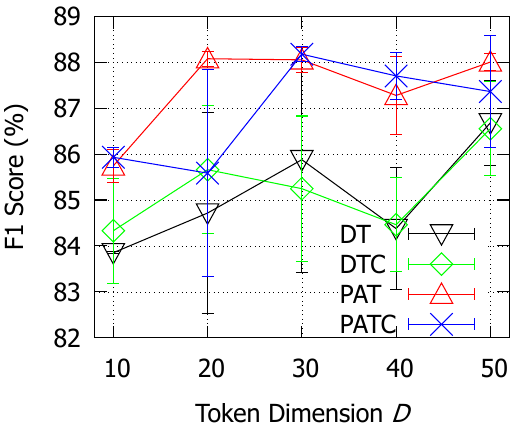}\label{fig:DwithNcompare5}}
        \caption{Effectiveness comparison of four Data Transformer methods with varying dimension $D$ and fixed number $N$. 
        }
        \label{fig:DwithNcompare}
    \end{figure}
\begin{figure}[t]
        \centering
        \subfloat[Restaurants]{\includegraphics[width=0.248\textwidth]{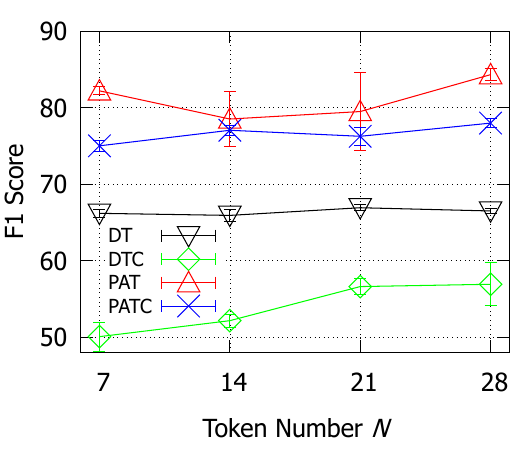}\label{fig:DfixedNcompare1}}
        \subfloat[Rayyan]{\includegraphics[width=0.248\textwidth]{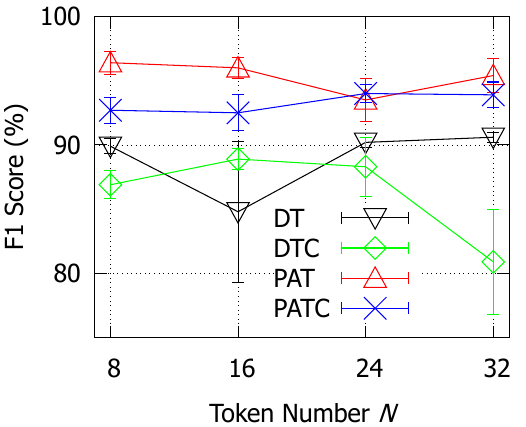}\label{fig:DfixedNcompare2}}
        \\
        \subfloat[Beers]{\includegraphics[width=0.248\textwidth]{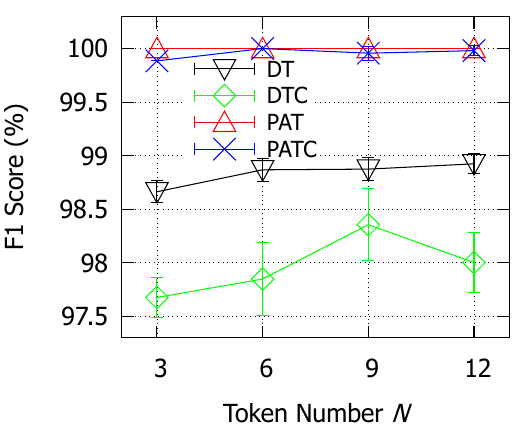}\label{fig:DfixedNcompare3}}
        \subfloat[Flights]{\includegraphics[width=0.248\textwidth]{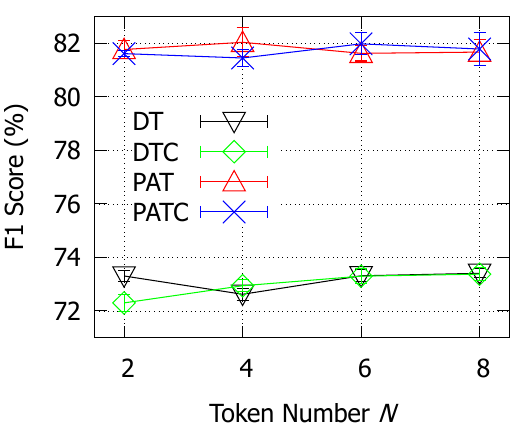}\label{fig:DfixedNcompare4}}
        \caption{ Effectiveness comparison of four Data Transformer methods with varying number $N$ and fixed dimension $D$. 
        }
        \label{fig:DfixedNcompare}
    \end{figure}

In Fig.\ref{fig:TypeCompare}, error detection works are compared on Rayyan and Flights across different error rates. PAT performs best results and outweighs 6 baselines on detecting syntactic and semantic errors in Rayyan, and syntactic errors in Flights. It can be ascribed to the learnable pattern module which is integrated into a single model. On the other hand, in Fig.\ref{fig:TypeCompare}(d), Flights with semantic errors  has similar and confused attributes (departure time, arrival time, etc.), and has few semantic error types (AD), which lead to that ZeroED and ETSB-RNN who have classifiers with few parameters learn better pattern features than PAT.

{\bf{Exp 2: Data Transformer performance evaluation.}}
In this experiment, we compare the F1 score of four Data Transformer methods (PAT, PATC, DT, DTC) for error detection. 
In Fig.~\ref{fig:DwithNcompare} and Fig.~\ref{fig:DfixedNcompare},
the token dimension $D$ and token number $N$ vary separately.

In Fig.~\ref{fig:DwithNcompare} and Fig.~\ref{fig:DfixedNcompare}, the accuracy of PAT and PATC is significantly higher than DT and DTC with a noticeable gap, for the PAT and PATC incorporate the learned attribute patterns. 
In Fig.~\ref{fig:DwithNcompare}(a,b,d) and Fig.~\ref{fig:DfixedNcompare}(a,b), the F1 scores of the PAT improve as the token dimension $D$ or token number $N$ increases separately.
This steady growth stems from the rough and mass tokenization in the QTA tokenizer to effectively accommodate long sequences in the above datasets (Restaurants, Rayyan, and HOSP-100K), where large tokens are generated by enlarging $D$ or $N$.

However, Fig.~\ref{fig:DwithNcompare}(c) and Fig.~\ref{fig:DfixedNcompare}(b) reveals gradual declines in F1 scores for certain methods once their optimal token dimension $D$ threshold is exceeded:
There are some declines in PAT from $D=36$, PATC and DTC both from $D=24$ in Fig.~\ref{fig:DwithNcompare}(c); 
PAT from $D=8$ and DTC from $D=16$ in Fig.~\ref{fig:DfixedNcompare}(b).
To account for the declining, most errors are spread over datasets with short or medium sequences, while the success rate of detecting these errors in short or medium sequences will be damaged when $D$ or $N$ increase excessively, as the token vectors generated by the QTA algorithm will contain more redundancy (zero-padding).

The four Data Transformer methods in Fig.~\ref{fig:DfixedNcompare}(c,d) show a stably invariable trend because full tokenization in the QTA tokenizer will provide delicate data tokens for short sequences datasets whenever scaling $D$ or $N$.
Consequently, we can draw that appropriate tokenization parameters ($D$, $N$) for the specific dataset can facilitate the model learning and improve accuracy performance.

\begin{table}[]
	\centering
	\caption{Computation Efficiency Comparison with State-of-the-Art Methods in Data Sets with Short, Medium, and Long Sequences}
	\label{tab_efficiency}
	\begin{tabular}{p{1.2cm}|p{0.68cm}p{0.63cm}|p{0.68cm}p{0.60cm}|p{0.68cm}p{0.63cm}}
		\toprule
		\multirow{2}{*}{Approach}                              & \multicolumn{2}{c|}{S Seqs DataSet}                                                                                          & \multicolumn{2}{c|}{M Seqs DataSet}                                                                                         & \multicolumn{2}{c}{L Seqs DataSet}                                                                                            \\ \cline{2-7} 
		& \begin{tabular}[c]{@{}l@{}}Params\\ ($\times10^6$)\end{tabular} & \begin{tabular}[c]{@{}l@{}}FLOPs\\ ($\times10^6$)\end{tabular} & \begin{tabular}[c]{@{}l@{}}Params\\ ($\times10^6$)\end{tabular} & \begin{tabular}[c]{@{}l@{}}FLOPs\\ ($\times10^6$)\end{tabular} & \begin{tabular}[c]{@{}l@{}}Params\\ ($\times10^6$)\end{tabular} & \begin{tabular}[c]{@{}l@{}}FLOPs\\ ($\times10^6$)\end{tabular} \\ \hline
		\begin{tabular}[c]{@{}l@{}}ETSB\\ -RNN\end{tabular}    & 0.06                                                            & 0.2                                                            & 0.08                                                            & 0.3                                                            & 0.08                                                            & 0.3                                                            \\ \hline
		\begin{tabular}[c]{@{}l@{}}HOLOD\\ -ETECT\end{tabular} & 8.3                                                             & 49.9                                                           & 13.1                                                            & 78.4                                                           & 14.6                                                            & 87.3                                                           \\ \hline
		ROTOM                                                  & 124.6                                                           & 99289.0                                                        & 124.6                                                           & 201296.0                                                       & 124.6                                                           & 319315.0                                                       \\ \hline
		DT                                                     & 6.1                                                             & 36.8                                                           & 82.4                                                            & 494.1                                                          & 120.0                                                           & 719.7                                                          \\
		DTC                                                    & 1.9                                                             & 11.5                                                           & 13.6                                                            & 81.8                                                           & 11.4                                                            & 68.3                                                           \\ \hline
		PAT                                                    & 15.4                                                            & 92.4                                                           & 231.7                                                           & 1390.4                                                         & 283.7                                                           & 1702.0                                                         \\
		PATC                                                   & 5.0                                                             & 29.9                                                           & 34.5                                                            & 206.8                                                          & 28.9                                                            & 173.7                                                          \\         
		\bottomrule
	\end{tabular}
\end{table}

{\bf{Exp 3: Computation quantity and efficiency evaluation.}}
Table~\ref{tab_efficiency} estimates the computational efficiency of 7 error detection methods including 4 Data Transformer models (PAT, PATC, DT, DTC models) and 3 types of EDT methods (single-model, multi-detector, and large-model approaches). 
We use two metrics, parameters and floating-point operations (FLOPs) on datasets with short, medium, and long sequences. 

There are two ways to calculate parameters and FLOPs.
For baselines, we assume that the counts in multiply-add operations (MACs) and the trainable parameters are approximately equal.
Then, we derive the FLOPs since it is twofold than MACs.
For Data Transformer methods, we divide the PAT model into three parts to count the FLOPs, including multiheaded self-attention, feed-forward and classification head. The FLOPs of all three parts are summed to obtain the total FLOPs.
Specifically, we incorporate all input tokens per sequence into FLOPs computing according to DeepMind's method~\cite{hoffmann2022training}.  
On account of one training step including forward pass and backward pass, the computation of the training procedure is defined as $C$$\approx$$6N_{\mathrm{tokens}}N_{\mathrm{params}}$ FLOPs per sequence~\cite{kaplan2020scaling}.

\begin{figure*}[htbp]
        \centering
        \includegraphics[width=1.0\textwidth]{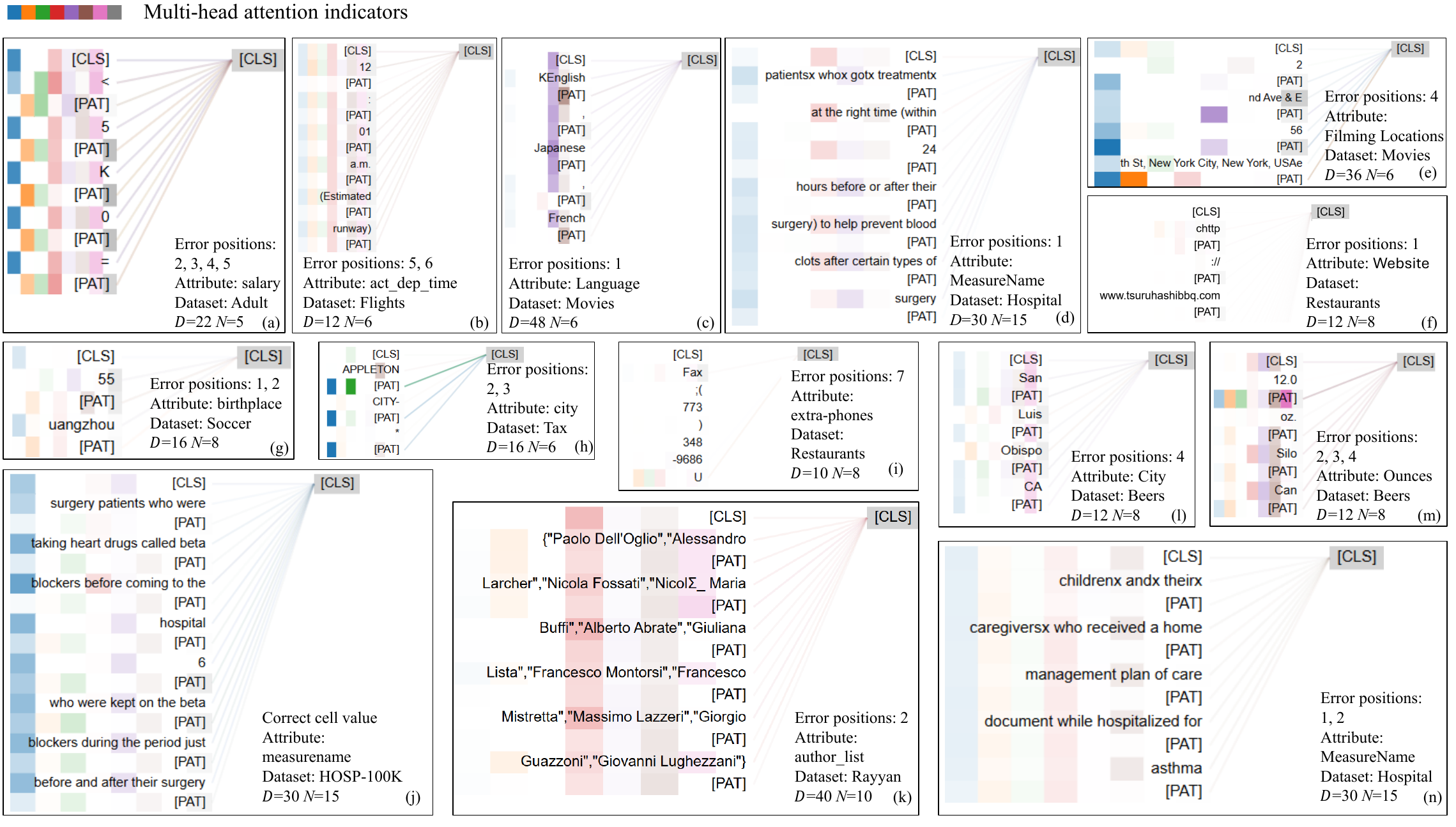}
        \caption{  Attention Visualization for Tabular Cell Instances.         }
        \label{attn_vis}
        \end{figure*} 

        \begin{figure}[!t]
        \centering
        \includegraphics[width=0.27\textwidth]{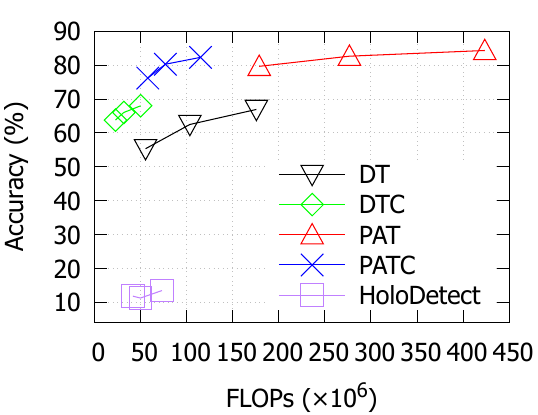}
        \caption{ Efficiency and Accuracy Measurement for Error Detection Methods.   }
        \label{tab_efficiency_accuracy_comparison}
\end{figure} 

As shown in Table~\ref{tab_efficiency},
The FLOPs of PAT and DT are more than twice that of PATC, DTC, where the increased computation is caused by incorporating the learned pattern tokens and is in exchange for improvement in accuracy.
Thus, PATC significantly outperforms PAT in efficiency while maintaining accuracy.
Nevertheless, PAT and PATC retain advantages over baselines:
As a single-model approach, ETSB-RNN has low computational cost but limited scalability. 
Possessed of multiple attribute-wise detectors, HoloDetect computes the FLOPs which are $C$$=$$|A|$ times that of a single detector model.
Since defining multi-level features as input, HoloDetect has more parameters and large cost of FLOPs for the long sequences.
As the single-model data error detection approach, PAT and PATC ensure simplicity and architectural integrity compared to HoloDetect.
Furthermore,
ROTOM, which uses Roberta, consumes large FLOPs that are computed under 3 sequence length levels (50, 200, 300)~\cite{calflops}.
PAT and PATC have fewer FLOPs than ROTOM (large-model approach), which demonstrates that PAT architecture can scale down the model size and meanwhile achieve even better performance.

To intuitively compare error detection methods,
Fig.~\ref{tab_efficiency_accuracy_comparison} reports FLOPs-F1Score line charts for efficiency and accuracy measurement. 
In Data Transformer methods (DT, DTC, PAT, and PATC), the lines show a trend of increasing accuracy when FLOPs increase in 3 scales. 
In terms of accuracy, our models of teh PAT series consistently outperform HoloDetect and models of the DT series in F1 score by a significant margin, under the conditions that the five models have narrow computations. 
Especially, PATC emphasizes efficiency and meanwhile preserves high accuracy when compared with PAT.

{\bf{Exp 4: Explainability of the pattern-perceptive Transformer.}}
By leveraging the input data, including a CLS token, data tokens and their corresponding pattern tokens, the multi-head attention (MSA) network learns features randomly distributed across the 8 parallel attention heads.
As illustrated in Fig.~\ref{attn_vis}, most attention maps are derived from the PAT model. Exceptions include Fig.~\ref{attn_vis}(c,e,f,k), which depict the PATC model, and Fig.~\ref{attn_vis}(i), which shows the DTC model. The error positions in cell examples are determined by dataset groundtruth.

In the MSA (8-head) network, different attention heads assume roles in data, pattern tokens or both. Specifically, the attention heads (e.g., heads 1 and 3 in Fig.~\ref{attn_vis}(b), head 3 in Fig.~\ref{attn_vis}(e), head 6 in Fig.~\ref{attn_vis}(d), and head 6 in Fig.~\ref{attn_vis}(n) with head indexed starting at 1) focus on extracting features from data tokens, while others (e.g., head 2 in Fig.~\ref{attn_vis}(a), head 1 in Fig.~\ref{attn_vis}(h), heads 3 and 6 in Fig.~\ref{attn_vis}(j), head 2 in Fig.~\ref{attn_vis}(m)) concentrate on learning from pattern tokens. What's more, some (e.g., head 5 in Fig.~\ref{attn_vis}(c), heads 4 and 6 in Fig.~\ref{attn_vis}(k)) are aimed to learn data and its pattern features.

For the error-contained tokens, two typical distributions of attention features are observed:
In the first case, high attention scores indicate a stronger focus on corresponding tokens.
\texttt{High-attention error tokens:} Error-contained tokens present dark regions (high attention scores) at data tokens (e.g., head 3 in Fig.~\ref{attn_vis}(e), head 2 in Fig.~\ref{attn_vis}(f), heads 2, 3 and 4 in Fig.~\ref{attn_vis}(i), heads 1, 3 and 5 in Fig.~\ref{attn_vis}(g), heads 4 and 5 in Fig.~\ref{attn_vis}(m)), their PAT pattern tokens (e.g., heads 1, 2, and 4 in Fig.~\ref{attn_vis}(e)), or both of data and PAT tokens (e.g., head 4 in Fig.~\ref{attn_vis}(f)). Whileas correct tokens are low attention scores with light colors.
For the second case, due to the symmetry of binary classification (clean and erroneous), an inversion of categorical feature representation occurs during training, leading to the counterintuitive low attention on error-contained tokens.
\texttt{Low-attention error tokens:} Error-contained tokens are low attention scores which display light or blank regions at either erroneous data tokens (e.g., head 6 in Fig.~\ref{attn_vis}(h), heads 2 and 4 in Fig.~\ref{attn_vis}(l)) or their PAT pattern tokens (e.g., heads 3 and 6 in Fig.~\ref{attn_vis}(l), head 3 in Fig.~\ref{attn_vis}(m)), whereas clean tokens show dark regions (high attention scores).
We can draw that 
some attention heads (or a single head) specialize in learning from error-contained tokens, and others focus on normal parts of the sequence.
The final MLP head module integrates all parallel attention heads collaboratively to make the final prediction.

Notably, MSA structures where the attention map in Fig.~\ref{attn_vis} is retrieved reside in the last (6th) encoder layer. It is found that the attention network in the deeper encoder layers in the PAT network can construct associations more clearly between the interleaved input tokens and output classification (CLS embedding). 
Moreover, the word-adaptive data tokens with the interleaved pattern tokens will also benefit the established relations, 
thereby accelerating the PAT’s error detection process.

{\bf{Exp 5: Ablation study of data embedding in PAT model.}}

In the PAT model,
the trainable linear projection (embedding layer) is not in Transformer encoders. 
To evaluate whether to use embedding layer or not will improve the detection performance, 
we present three variations of data embedding in our PAT method and conduct the ablation study on four datasets. 
(1) Unicode character mapping is applied in the QTA tokenizer to verify the effectiveness of character embedding. This way is default setting.
(2) MLP (Linear layers) is appended to the PAT network to validate the effectiveness of the embedding layer.
(3) Word2Vec model (fastText) is used in the QTA tokenizer to verify the effectiveness of word embedding.

Based on Table~\ref{tab_ablation_embedding}, 
First, our primary finding is that PAT with the character embedding strategy is more effective on accuracy of the EDT task than the other versions.
In terms of accuracy, PAT with character embedding outweighs the responding versions equipped with the MLP layers and fastText in Movies, Restaurants and Flights datasets.  
PATC with character embedding outweighs all the other versions in Rayyan.
According to the ablation study, 
first, PAT network without embedding layer will retain the superficial format and elementary sequence characteristics, since our error detection model focuses more on violations related with character composition and format i.e. syntactic errors. 
Second, 
although dropping the embedding layer and word embedding strategies, 
PAT framework with the character index mapping can provide eligible demands for detecting semantic-oriented errors, i.e. AD.
It can be attributed to that dropping the embedding layer in the PAT model will put off constructing the semantic relations among input tokens in the subsequent deeper PAT network.

\begin{table}[]
        \centering
        \caption{Ablation Study of Data Embedding}
        \label{tab_ablation_embedding}
        \begin{tabular}{p{0.16cm}p{0.16cm}|cp{0.30cm}c|cp{0.30cm}c}
        \toprule
        \multicolumn{2}{c|}{Methods}                                                                       &               & PAT           &                                                                                                    &               & PATC          &                                                                                                    \\ \hline
        \multicolumn{2}{c|}{\begin{tabular}[c]{@{}c@{}}Data\\Embedding\end{tabular}}                     & \begin{tabular}[c]{@{}c@{}}Char\\Map \end{tabular}       & MLP           & \begin{tabular}[c]{@{}c@{}}fastText\\~\cite{bojanowski2017enriching} \end{tabular} & \begin{tabular}[c]{@{}c@{}}Char\\Map \end{tabular}       & MLP           & \begin{tabular}[c]{@{}c@{}}fastText\\~\cite{bojanowski2017enriching} \end{tabular} \\ \hline
        \multicolumn{1}{l|}{\multirow{3}{*}{Movies}}                                                  & P  & 96.7          & 93.5          & 98.9                                                                                               & 95.8          & 94.1          & \textbf{99.3}                                                                                      \\
        \multicolumn{1}{l|}{}                                                                         & R  & \textbf{62.9} & 54.5          & 27.3                                                                                               & 61.3          & 59.0          & 27.7                                                                                               \\
        \multicolumn{1}{l|}{}                                                                         & F1 & \textbf{76.2} & 68.7          & 42.7                                                                                               & 74.7          & 72.6          & 43.4                                                                                               \\ \hline
        \multicolumn{1}{l|}{\multirow{3}{*}{Rayyan}}                                                  & P  & 95.3          & 94.1          & 90.3                                                                                               & \textbf{96.9} & 93.2          & 91.1                                                                                               \\
        \multicolumn{1}{l|}{}                                                                         & R  & 95.3          & \textbf{97.0} & 87.5                                                                                               & 96.1          & 94.4          & 85.6                                                                                               \\
        \multicolumn{1}{l|}{}                                                                         & F1 & 95.3          & 95.5          & 88.6                                                                                               & \textbf{96.5} & 93.7          & 88.3                                                                                               \\ \hline
        \multicolumn{1}{l|}{\multirow{3}{*}{\begin{tabular}[c]{@{}l@{}}Restau\\ -rants\end{tabular}}} & P  & \textbf{97.1} & 96.6          & 95.5                                                                                               & 96.7          & 96.3          & 94.1                                                                                               \\
        \multicolumn{1}{l|}{}                                                                         & R  & \textbf{74.8} & 69.6          & 45.3                                                                                               & 71.9          & 67.1          & 42.7                                                                                               \\
        \multicolumn{1}{l|}{}                                                                         & F1 & \textbf{84.5} & 80.8          & 61.4                                                                                               & 82.5          & 79.0          & 58.8                                                                                               \\ \hline
        \multicolumn{1}{l|}{\multirow{3}{*}{Flights}}                                                 & P  & 73.6          & 76.9          & 74.7                                                                                               & 75.8          & \textbf{79.1} & 70.9                                                                                               \\
        \multicolumn{1}{l|}{}                                                                         & R  & \textbf{93.3} & 87.8          & 91.7                                                                                               & 90.0          & 85.0          & 91.9                                                                                               \\
        \multicolumn{1}{l|}{}                                                                         & F1 & \textbf{82.0} & 81.7          & \textbf{82.0}                                                                                      & \textbf{82.0} & 81.7          & 80.0                                                                                               \\ \bottomrule
        \end{tabular}
        \end{table}

\section{CONCLUSION}
We propose an attribute-specific pattern-perceptive Transformer framework for the error detection of relational databases to overcome the accuracy and efficiency trade-off.
The data tokens generated from the QTA tokenizer and learned attribute-wise pattern tokens are jointly devoted to facilitating PAT model training, thereby putting distinguishable characteristics which is unique to each attribute domain and shared data features isolated and storing them in two types of parameters (data-specific part and pattern part) in PAT network. PAT is also supported for other ML tasks which have multi-patterns like data.
Experimental results verify that we evaluate model scalability with attribute values of long sequences and with large datasets with numerous tuples. 
In the future, we intend to solve more semantic errors.
The explainability mechanism of the error detection can be further developed and utilized for subsequent data repair and assessment.

\section{ACKNOWLEDGMENTS}
This work was supported in part by National Natural Science Foundation of China under Grants  62402135, U21A20513, Taishan Scholars Program of Shandong Province grant under Grant tsqn202211091, Shandong Provincial Natural Science Foundation under Grant ZR2023QF059.


\bibliographystyle{ACM-Reference-Format}
\bibliography{sample}

\end{document}